# From Flash to Crater: Morphological and Spectral Analysis of the Brightest Lunar Impact on 11 September 2013 using LRO Data

J. L. Rizos[1], L. M. Lara[1], J. L. Ortiz[1], J. M. Madiedo[1]

1 - Instituto de Astrofísica de Andalucía, CSIC, Apdo. 3004, Camino Bajo de Huétor 50, E-18080 Granada, Spain

## Abstract

We present a comprehensive morphological and spectrophotometric analysis of the lunar impact that occurred on September 11, 2013, based on pre- and post-event observations by the Lunar Reconnaissance Orbiter (LRO). The crater formed exhibits a rim-to-rim diameter of 35 ± 0.7 m, a depth of 4.9 ± 0.4 m, and an ejecta blanket extending over 2 km with an area of approximately $7 \times 10^5$ m². The ejecta shows a pronounced asymmetry and, assuming uniform distribution, an average thickness limit of ~2 mm. Spectral analysis using WAC images reveals a consistent reddening of the central ejecta region, with an average 16.54% increase in spectral slope between 321 nm and 643 nm, marking the first reported detection of color changes resulting from a lunar impact. We evaluated several scaling laws and found that the Gault formulation most accurately reproduces the observed crater size. Furthermore, luminous efficiency values below $\eta = 2 \times 10^{-3}$ and higher projectile densities are most consistent with the morphology of the ejecta. The impact direction inferred from this pattern is not compatible with the radiant of the September ε-Perseids stream. Moreover, an independent probability analysis yields a greater than 96% likelihood that the event was caused by a sporadic meteoroid. Our results also demonstrate the potential of WAC imagery for the automated detection of new lunar craters, which can improve statistical estimates of the current impact flux. This methodology offers a powerful complement to high-resolution imaging, with important implications for both lunar safety and planetary defense.

## 1. Introduction

Interplanetary matter mostly originate from comets and asteroids (e.g. Nesvorny et al., 2010), which act as the main reservoirs of dust and small debris in the Solar System. Beyond these dominant sources, a minor but identifiable fraction of the material originates from planetary bodies such as Mars (e.g., SNC meteorites; Treiman et al., 2000) and it has been speculated that even natural satellites like Phobos or Deimos may contribute as well (Wiegert and Galiazzo, 2017). A substantial fraction of this material consists of particles ranging from approximately 10 micrometers to 1 meter in size, which are gravitationally bound to the Sun and classified as meteoroids (Rubin and Grossman, 2010).

Most meteoroids, once detached from their parent bodies, initially travel through the Solar System in orbits very similar to those of their sources. Particles belonging to meteoroid streams tend to preserve such orbits, having typically separated from their progenitors less than a few thousand years ago. In contrast, sporadic meteoroids were ejected tens of thousands of years ago—or even

earlier—and their orbits have since undergone substantial alterations, making them appear randomly in both time and location (Jenniskens, 2006; Jopek and Williams, 2013). Both types of meteoroids represent a continuous flux of material that constantly exchange and supply material to the planets and moons. For example, Earth receives from 5 to 270 tones per day of interplanetary material (e.g. Plane, 2012). However, most of the larger fragments of material disintegrate in the atmosphere before reaching the ground, occasionally resulting in spectacular events such as fireball entries.

An analogue meteoroid bombardment also occurs on the Moon. However, unlike Earth, the Moon lacks a substantial atmosphere, allowing meteoroids to reach its surface and produce transient luminous phenomena—impact flashes—that can be observed from Earth. These hypervelocity collisions generate brief bursts of optical emission, primarily due to the formation of high-temperature plasma and, potentially, to the thermal emission from silicate-rich ejecta droplets as they cool (Yanagisawa and Kisaichi, 2002). In such high-speed impacts, the projectile is completely destroyed upon contact (Melosh, 1989). This continuous bombardment provides a unique opportunity to investigate the properties of interplanetary matter, as well as the physical characteristics of the lunar surface. Moreover, understanding the frequency and consequences of meteoroid impacts on the Moon is becoming increasingly relevant as humanity prepares for a new era of sustained lunar exploration. Over the next decade, multiple space agencies and commercial entities plan to deploy crewed missions, build permanent infrastructures, or conduct long-term surface operations on the Moon (e.g., NASA’s Artemis program, ESA’s Moonlight initiative, or China’s International Lunar Research Station). In such a context, even centimeter-sized meteoroids pose a potential hazard to astronauts, habitats, and surface assets. Therefore, characterizing this lunar impact environment is thus not only a matter of scientific interest but also of practical importance for ensuring the safety and sustainability of future lunar activities.

Moreover, the risk is not limited to lunar infrastructure but may also extend directly to Earth. As reported by Wiegert et al., (2025), asteroid 2024 $YR_4$ has a 4% probability of impacting the Moon on 22 December 2032 at a velocity of approximately 13 km/s. Such an impact could eject a fraction of lunar material beyond the Moon’s escape velocity. Melosh (1985) demonstrated that these impacts can eject around 0.01% of surface material at speeds exceeding escape velocity—most of which consists of micron- to millimeter-sized particles (Housen and Holsapple 2011). Depending on the impact location, up to 10% of this material could be gravitationally captured by Earth within a few days, potentially causing a substantial increase in the meteoroid flux in near-Earth space (Wiegert et al., 2025). This influx of lunar ejecta could expose satellites in low Earth orbit to an impact environment equivalent to several years—or even a decade—of background meteoroid flux. Moreover, a fraction of these particles may remain in orbit for extended periods, posing long-term risks to spacecraft operations. For all these reasons, understanding the consequences of lunar impacts—and being able to predict the resulting crater dimensions, as well as the volume and properties of the ejected material—is crucial for planetary defense, satellite risk assessment, and the near-term future of space exploration.

Systematic monitoring of these events using telescopic observations with CCD cameras began in the late 1990s (Ortiz et al., 1999), and since then, numerous flashes have been recorded both during major meteor showers and from sporadic meteoroids (e.g. Ortiz et al., 2000; Yanagisawa and Kisaichi, 2002; Ortiz et al., 2002; Yanagisawa et al., 2006; Suggs et al., 2014; Liakos et al., 2020; Yanagisawa et al., 2025). This effort was significantly advanced in 2009 with the launch of the Lunar Reconnaissance Orbiter (LRO) (Chin et al., 2007), which has been orbiting the Moon ever since, completing approximately 12 orbits per day in a near-polar, low-altitude trajectory. LRO carries a suite of instruments designed to study the lunar surface at high spatial and spectral resolution, enabling the identification and characterization of fresh impact craters. Among other instruments, LRO carries the Narrow Angle Camera (NAC), which provides high-resolution imaging in the visible range (panchromatic, 400-750 nm), and the Wide-Angle Camera (WAC), equipped with multiple color filters, including two in the near-ultraviolet and five in the visible range, with central wavelengths at 321, 360, 415, 566, 604, 643, and 689 nm, that enable multispectral observations. These two imaging systems are part of the Lunar Reconnaissance Orbiter Camera (LROC) suite (Robinson et al., 2010). The NAC provides a high spatial resolution of up to 0.5 meters per pixel, making it ideal for detailed morphological analyses of the lunar surface. In contrast, the WAC captures lower-resolution images across seven distinct spectral bands ranging from 320 to 690 nm, with a nominal resolution of approximately 400 meters per pixel in the ultraviolet and 100 meters per pixel in the visible. This multi-spectral capability of the WAC is crucial for studying compositional variations on the Moon.

The brightest and most energetic impact flash recorded to date was detected on 2013 September 11 at 20:07:28.68 ± 0.01 UTC (Madiedo et al., 2014). The event produced a flash with a peak visual magnitude of 2.9 ± 0.2 and the authors could determine its selenographic coordinates with high precision (Lat.: 17.2° ± 0.2° S, Lon.: 20.5° ± 0.2° W), reporting a total duration of 8.3 seconds and a kinetic energy equivalent to several tens of tons of TNT. Assuming a luminous efficiency of $\eta = 2 \times 10^{-3}$ (i. e. the fraction of the projectile's kinetic energy that is converted into light during the impact), they estimated the impactor's mass to be 450 ± 75 kg, with a diameter between 0.61 – 1.42 meters (<1.5 $m^3$) and predicted a crater diameter between 47 and 56 meters by assuming an impact angle of 45°.

In this work, we aim to characterize the morphological and spectral variations resulting from this impact by taking advantage of the unique opportunity provided by LRO to analyze pre- and post-impact observations of the affected region. On the one hand, the high-resolution NAC images offer detailed morphological information essential for understanding the consequences of this impact. On the other hand, the availability of color filters enables the investigation of potential compositional changes and represents a promising step forward in the scientific exploitation of LRO data; while ground-based flash observations face limitations due to lunar phases, daytime periods, or the meteorological conditions of their terrestrial location, WAC observations provide a consistent imaging cadence generating a global map every lunar rotation. This enables the search for lunar impacts not recorded by ground-based observations, based on color variations that may extend farther from the impact site than the features visible in high-resolution NAC images.

Although crater detection algorithms based on high-resolution NAC image comparison has been developed, their utility for identifying new impacts is limited (e.g. Sheward et al., 2022). This is primarily due to the NAC's long imaging cadence to cover the same region (typically of several years) and the significant differences in scenes caused by variations in viewing geometry such as the incidence angle, which alter the length and orientation of shadows and complicate automated detection. By contrast, analyzing color changes—less sensitive to viewing conditions and with a shorter revisit time—may prove more effective for the future detection of meteoroid impacts. Given that this study focuses on the largest impact recorded to date, the results presented here provide an upper bound on the expected changes associated with such events and may help guide analytical strategies aimed at maximizing the scientific return from LRO color data.

In Section 2, we describe the datasets used in this study and the preprocessing steps applied. The results are presented in Section 3, where we analyze both the morphological features of the crater and its ejecta, as well as the associated color changes. Section 4 is devoted to the discussion, including the likely origin of the impactor and the application of crater scaling laws. Finally, our main conclusions are summarized in Section 5.

## 2. Data Preparation

To locate the images used in this study, we employed the web interface for map-related products developed by ACT-REACT QuickMap[1]. This platform offers a search engine that becomes accessible once a region of interest is defined using the 'Draw and Query' tool, enabling users to retrieve all LROC images of that area stored in the database. We verify the availability of both all NAC and WAC images covering the period from 2009 to February 2025 at the time of writing. Moreover, the search engine enables filtering based on average values of phase, emission and incidence angles[2], resolution, or even acquisition date. Subsequently, we download from the LROC data portal[3] the uncalibrated Experiment Data Record (EDR) images. The following sections describe the calibration and preprocessing procedures applied to each instrument dataset:

*NAC images*: The raw images (IMG format) were first converted into ISIS (Integrated Software for Imagers and Spectrometers[4]) cube format. Next, geometric information from the LRO mission SPICE kernels was added to each image using the spiceinit function from ISIS, which computes the necessary ancillary data such as spacecraft position, pointing, and solar geometry. Then, images were converted to I/F[5] using the processing lronaccal pipeline[6], which is available within the ISIS framework. The images were then map-projected using an equirectangular projection (Eq. 1).

$$\begin{cases} x = R(\lambda - \lambda_0)\cos(\phi_1) \\ y = R(\phi - \phi_1) \end{cases} \quad (1)$$

where $\lambda$ is the longitude of the point, $\lambda_0$ is the central longitude, $\phi$ is the latitude of the point, $\phi_1$ is the central latitude, and $R$ is the radius of the Moon.

For comparison purposes, we selected pre- and post-impact images with similar phase, emission, and incidence. We found a residual pointing error after georeferencing the images, resulting in a displacement of several pixels. To correct this, we converted the images to TIFF format and used

[1] https://quickmap.lroc.im-ldi.com
[2] Where *phase angle* is defined between the Sun and the observer (i.e., LROC), *emission angle* is measured from the surface normal to the observer, and *incidence angle* from the surface normal to the solar vector.
[3] https://lroc.im-ldi.com
[4] ISIS is a specialized software package developed by the USGS for the processing of planetary remote sensing data: https://isis.astrogeology.usgs.gov/8.1.0/
[5] I/F, also referred to in the literature as the radiance factor (RADF), is the ratio of the bidirectional reflectance of a surface to that of a perfectly diffuse surface illuminated at $i$ = 0°.
[6] https://isis.astrogeology.usgs.gov/9.0.0/Application/presentation/Tabbed/lronaccal/lronaccal.html

QGIS[7]. This software provides a user-friendly interface that allows the co-registration of image pairs by identifying common tie points, with one image taken as the reference. For each dataset, we selected more at least 30 control points and applied the QGIS *Thin Plate Spline* transformation. A visual inspection using image blinking confirmed that the alignment achieved was accurate, with no perceptible differences to the human eye. The NAC images used in this work are listed in Table 1.

**Table 1. List of NAC images used in this study, acquired before and after the 11 September 2013 lunar impact. The table includes the image ID, acquisition date, spatial resolution (in meters per pixel), and the mean emission, incidence, and phase angles from each frame.**

| ID | Date | Resolution (m/pixel) | Emission (°) | Incidence (°) | Phase (°) |
|---|---|---|---|---|---|
| Pre-impact | | | | | |
| M124707982RE | 2010-03-31T20:51:57 | 0.50 | 1.14 | 18.99 | 18.8 |
| M193067490LE | 2012-05-31T01:37:02 | 3.15 | 51.73 | 76.63 | 25.62 |
| M1096630203RE | 2012-07-11T07:15:35 | 0.79 | 1.16 | 68.52 | 67.37 |
| M1119014742LE | 2013-03-27T09:11:14 | 0.82 | 1.73 | 23.66 | 24.83 |
| Post-impact | | | | | |
| M1154349871LE | 2014-05-10T08:30:03 | 0.96 | 3.84 | 70.42 | 74.23 |
| M1276711033RE | 2018-03-26T13:42:45 | 1.56 | 1.16 | 85.02 | 83.87 |
| M1284953808LE | 2018-06-29T23:22:20 | 0.68 | 1.71 | 16.50 | 16.76 |
| M1293170489RE | 2018-10-03T01:47:01 | 1.65 | 1.16 | 81.57 | 80.42 |
| M1320214670RE | 2019-08-12T02:03:22 | 0.64 | 1.15 | 65.58 | 64.43 |
| M1371894473LE | 2021-04-01T05:33:25 | 0.54 | 2.57 | 28.85 | 30.99 |
| M1417641871LE | 2022-09-12T17:10:03 | 0.89 | 1.73 | 18.68 | 19.17 |
| M1435252023RE | 2023-04-04T12:52:35 | 0.62 | 1.15 | 38.82 | 37.71 |
| M1441095458LE | 2023-06-11T04:03:10 | 2.06 | 1.74 | 71.06 | 72.79 |
| M1445785486RE | 2023-08-04T10:50:18 | 1.02 | 1.17 | 22.71 | 22.01 |
| M1465725267LE | 2024-03-22T05:39:59 | 0.76 | 1.71 | 57.48 | 59.19 |
| M1471562612RE | 2024-05-28T19:09:04 | 0.89 | 1.16 | 51.01 | 49.88 |

*WAC images*. As for the WAC dataset, raw images in IMG format were first converted into ISIS cube format, georeferenced, and calibrated to I/F using the lrowaccal pipeline[8], which is also available within ISIS. Due to the acquisition strategy of the WAC, images are captured in short framelets of 14 lines per integration. When converting images into cubes, the software generates two separate cube files per band: one containing the “even” framelets and another containing the “odd” ones. This division reflects the acquisition pattern of the instrument, which captures alternating framelets in an interleaved sequence. To reconstruct a continuous image, these cubes were mosaicked using appropriate tools within ISIS. Next, the images were map-projected using the equirectangular projection (Eq. 1). The WAC images used in this analysis are summarized in Table 2.

[7] https://qgis.org/

[8] https://isis.astrogeology.usgs.gov/9.0.0/Application/presentation/Tabbed/lrowaccal/lrowaccal.html

**Table 2. List of WAC images used in this work, acquired before and after the 11 September 2013 lunar impact. The table includes the image ID, acquisition date, spatial resolution (in meters per pixel), and the mean emission, incidence, and phase angles from each frame.**

| ID | Date | Resolution (m/pixel) | Emission (°) | Incidence (°) | Phase (°) |
|---|---|---|---|---|---|
| Pre-impact | | | | | |
| M109365846CE | 2009-10-05T07:09:38 | 68.86 | 1.13 | 13.89 | 12.86 |
| M155369258CE | 2011-03-21T17:53:10 | 60.44 | 1.14 | 14.32 | 15.45 |
| M170695002CE | 2011-09-15T03:02:14 | 58.48 | 1.14 | 14.07 | 15.04 |
| Post-impact | | | | | |
| M1223802480CE | 2016-07-22T04:53:32 | 76.83 | 1.15 | 13.54 | 14.40 |
| M1223809425CE | 2016-07-22T06:49:17 | 77.10 | 1.15 | 12.52 | 13.28 |
| M1315513324CE | 2019-06-18T16:07:36 | 77.70 | 1.15 | 13.24 | 12.72 |

*Photometric corrections*. The first step before applying any correction or normalization to the images is to generate photometric backplanes, which include phase, emission and incidence maps. For NAC images, we generate backplanes using the 5.00 m/pixel DTM[9] produced by the LROC team for this region (Henriksen et al., 2017). In the case of NAC, we compute pre-/post-impact image ratios, which effectively cancel out most of the effects related to viewing geometry. Only a disk function was necessary to account for small differences in incidence and emission angles. As disk function we used the Akimov model (Akimov, 1988a, 1988b; Akimov et al., 2000, 1999; Shkuratov et al., 2011) (Eq. 2), which has been shown to be one of the most accurate empirical descriptions of lunar photometric behavior (Korokhin et al., 2018), providing a significantly good fit to the observed scattering characteristics of the lunar surface:

$$D(\alpha, \beta, \gamma) = \cos\left(\frac{\alpha}{2}\right)(\cos\beta)^{\nu\alpha/(\pi-\alpha)}\cos\left[\frac{(\gamma-\alpha/2)\pi}{\pi-\alpha}\right](\cos\gamma)^{-1} \qquad (2)$$

where $\beta$ and $\gamma$ are the photometric latitude and longitude, defined as:

$$\gamma = \arctan\left(\frac{\cos(i)-\cos(e)\cdot\cos(\alpha)}{\cos(e)\cdot\sin(\alpha)}\right) \qquad \beta = \arccos\left(\frac{\cos(e)}{\cos(\gamma)}\right)$$

[9] https://data.lroc.im-ldi.com/lroc/view_rdr/NAC_DTM_SEPTIMPACT

with $\alpha$, $e$ and $i$ being the phase, emission and incidence angles, respectively. The coefficient $\nu$, which characterizes sub-pixel roughness, was set to 0.43—an intermediate value between 0.34 (maria) and 0.52 (highlands) (Korokhin et al., 2018).

For WAC images, since their resolution is ~100 m/pixel, we generate the photometric backplanes from a near-global lunar topographic digital terrain model produced by Scholten et al., (2012), derived from stereo observations by WAC with a nominal spatial resolution of 100 m/pixel. Given that WAC images can extend up to 50 km horizontally, it was necessary to apply not only a disk but also a phase correction. To do so, we followed the approach outlined by Korokhin et al., (2018) for the photometric correction of WAC data. This method models the reflectance as the product of a disk function (Akimov, Eq. 2) and a phase function. For the latter, we adopted the empirical formulation proposed by Korokhin et al., (2007) (Eq. 3), which is valid for phase angles between 6° and 120°:

$$f(\alpha) = A_{n1}\mathrm{e}^{-\rho\alpha} + A_{n2}\mathrm{e}^{-0.7\alpha} \qquad (3)$$

where $\alpha$ is the phase angle in radians, and the parameters $A_{n1}$, $A_{n2}$, and $\rho$ were retrieved from Table 3 of Korokhin (2007), selecting the values corresponding to the closest effective wavelength for each case.

## 3. Results

### *3.1 Crater Dimensions and Ejecta Morphology*

We used several projected images taken at different incidence angles to better define the crater's edges. The structure appears to be nearly circular, with a horizontal (west to east) rim crest to rim crest diameter of 37 ± 1 m and a vertical (south to north) of 33 ± 1 m (Fig. 1). This yields an average diameter $D$ of 35 ± 0.7 m, corresponding to a total surface area of 962 ± 39 m². Note that, according to Melosh (1989), rim-to-rim diameters are approximately 25% larger than the apparent diameters. Applying this relation gives a value of $D_{ap}$ = 28 ± 0.6 m.

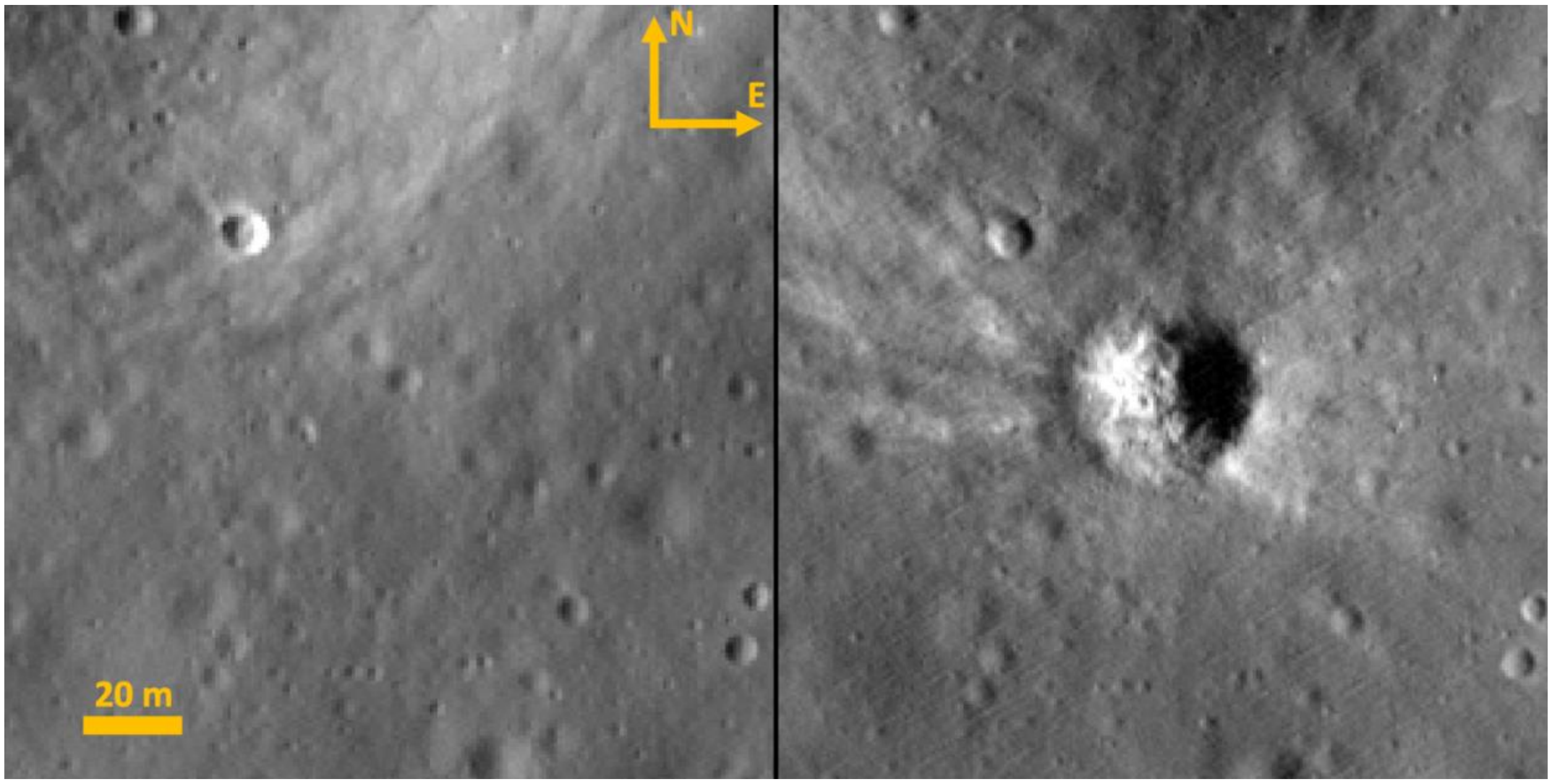

**Fig 1. Left panel: M1096630203RE NAC nadir image acquired at an incidence angle of 68.52° and phase angle of 67.37°, with a spatial resolution of 0.79 m/pixel on 2012-07-11. Right panel: Same region but with the newly formed crater as seen in NAC nadir image M1320214670RE taken on 2019-08-12 at an incidence angle of 65.58° and phase angle of 64.43°, very similar to those of the pre-impact image, with a resolution of 0.64 m/pixel. The center of the new crater is located at 17.16°S and 20.41°W.**

To estimate the depth, we first used the 5.00 m/pixel DTM. In Fig. 2 the shape model is shown on the left, and the elevation profile along the drawn trace is shown on the right. To derive a depth value, we fitted a straight line between the outer rims of the crater and measured the vertical distance from this line to the minimum point (pixel 7), obtaining a depth of 4.3 meters. However, according to Henriksen et al., (2017), the vertical accuracy can be up to several times the pixel scale, so the associated uncertainty is on the order of a few meters.

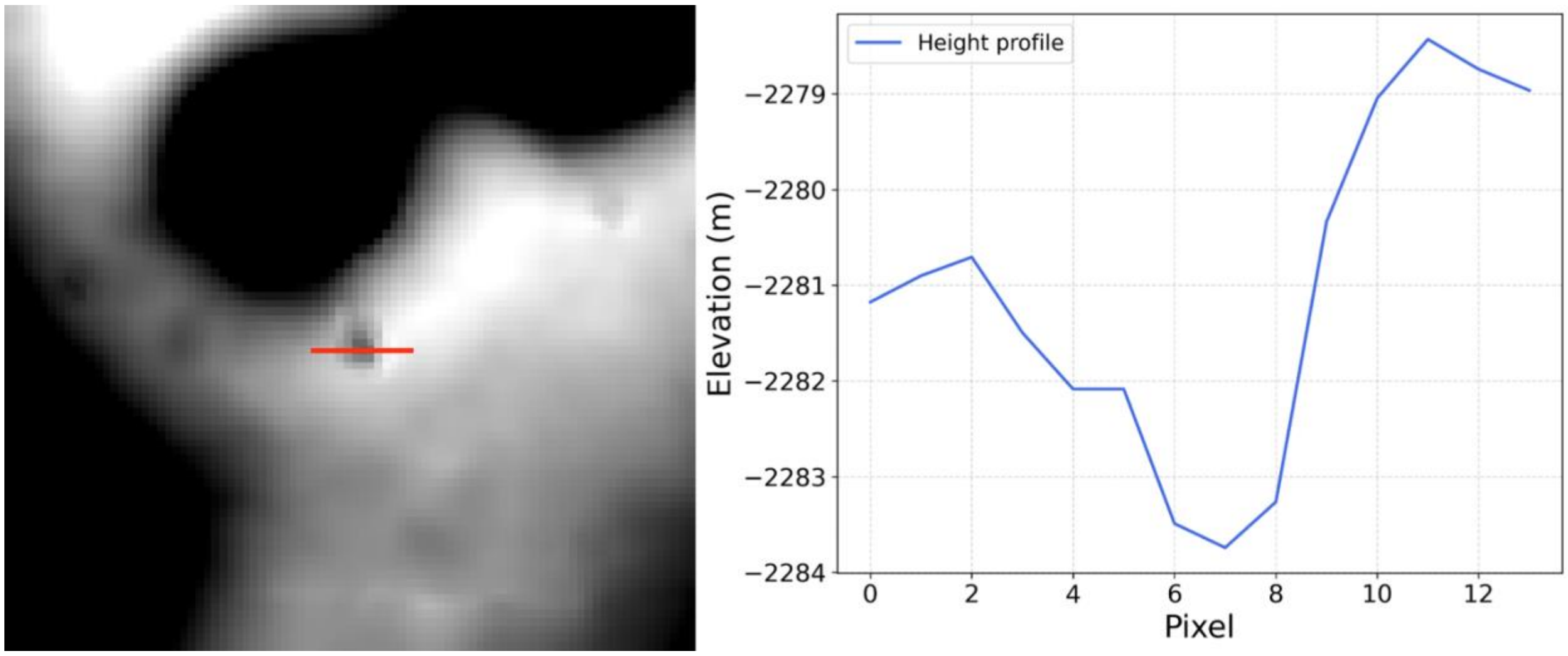

**Fig 2. Left: 5 m/pixel DTM centered on the new crater analyzed in this work. Right: elevation profile (referenced to the lunar mean radius), extracted along the red line crossing the crater in the left panel. The crater depth is 4.3 meters in this DTM.**

To provide an independent measurement and increase confidence in our depth estimate, we analyzed shadow lengths following the scheme illustrated in Fig. A1. In this case, since the images were acquired at nadir, the incidence angle ($i$) of the region and the shadow length ($L$) are trigonometrically related to the depth measured from the crater rim crest to the shadow terminus ($d'$) as $d' \sim L \tan(90 - i)$.

We applied this method to the set of post-impact images taken at different incidence angles (Table 1), computing the subsolar direction for each case and converting the shadow length accordingly (see example in Fig. A2). The maximum value of $d'$ was obtained from NAC image M1154349871LE, with the crater observed at an incidence angle of 69.6°. It was measured near the crater center (bottom) and corresponds to $d' = 6.3 \pm 0.4$ meters. According to Stopar et al., (2017), extremely fresh craters show a rim crest height to crater diameter ratio (*h/D*) of 0.04 (rim crest height measured above the surrounding terrain baseline, $h$ in Fig. A1). In this case, that implies an expected rim crest height $h$ of 1.4 m, therefore, we can establish a lower limit for $d$ of 4.9 ± 0.4 m. Extremely fresh craters typically have a depth-to-diameter (*d/D*) ratio between 0.11 and 0.17 (Stopar et al., 2017), so using the 35 m diameter calculated previously, it implies an expected depth between 3.85 and 5.95 meters, in agreement with our estimate.

We then analyzed the morphological distribution of the ejecta. As a preliminary step, we visually examined how viewing geometry affects its visibility. We found that the ejecta is more evident under lower incidence angles (Fig. A3)—which in LRO nadir observations also correspond to lower phase angles, since both angles are coupled. Therefore, to analyze the morphology of the ejecta, we searched among all pre- and post-impact image pairs for the combination with the highest spatial resolution and lowest incidence angle. We identify the best pair as the ratio M1417641871LE/M1119014742LE; it provides an average spatial resolution of 0.89 m/pixel, cover the ejecta extent completely, and the incidence angle values are as low as 18.68° and 23.66°, respectively. The resulting image ratio after disk photometric corrections is shown in Fig. 3.

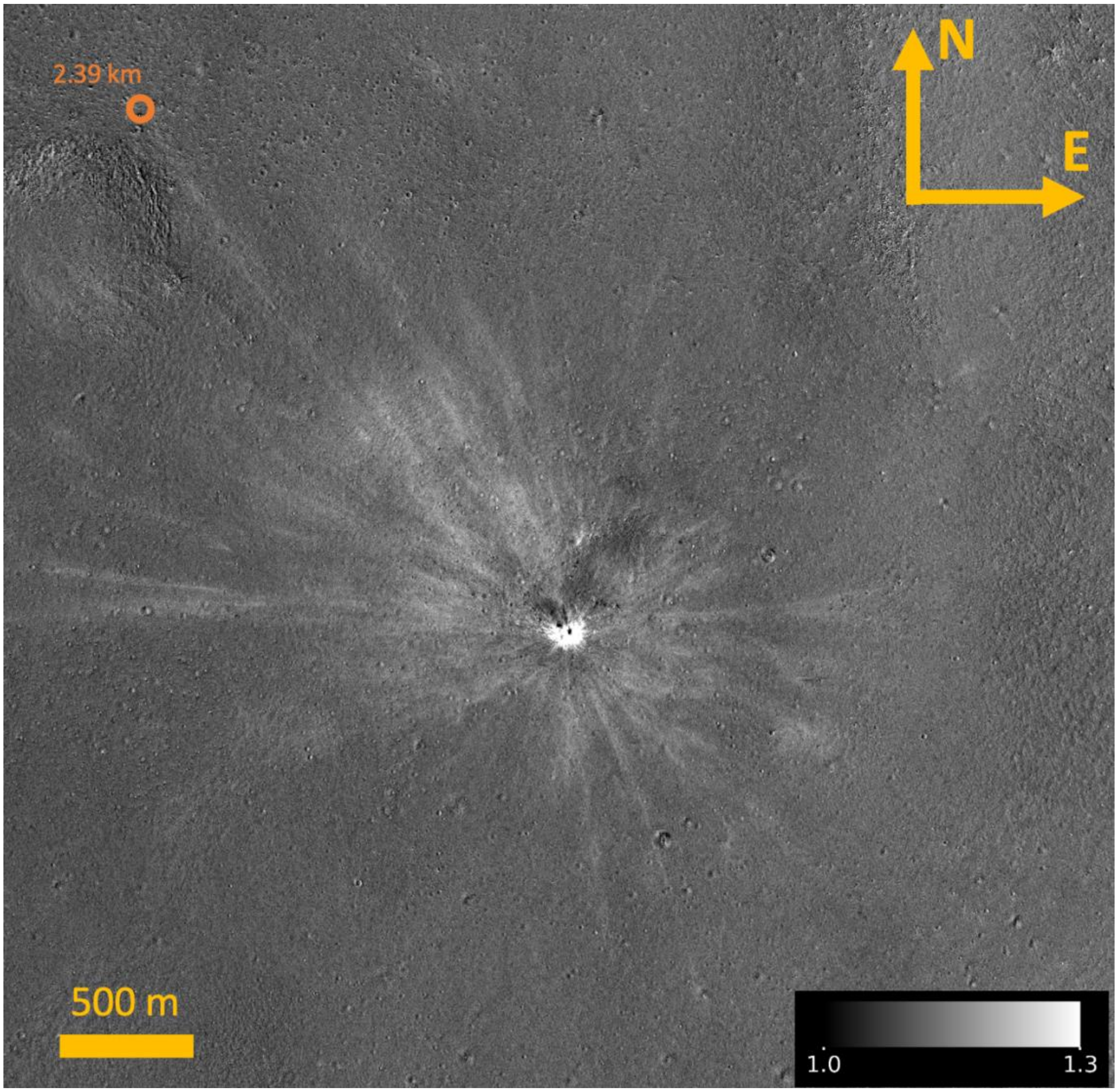

**Fig. 3. Ratio image M1417641871LE / M1119014742LE after applying Akimov disk photometric correction to both images. The ejecta appears brighter than the background. The most distant ejecta is located 2.39 km from the crater (orange circle in the upper left corner).**

This procedure reveals ejecta extending over distances larger than 2 km. To analyze its morphology more effectively, it was necessary to isolate the ejecta as much as possible. For this purpose, we applied the *k-means* clustering algorithm, which has been previously validated on similar datasets from the DAWN and OSIRIS-REx missions at Ceres and Bennu, respectively (Rizos et al., 2019, 2021). We first decomposed the data into clusters by testing a range of cluster numbers ($k$ = 2 to 13) and visually evaluating the resulting segmentations. We found that using $k$ = 3 yielded two clusters covering the ejecta and a third corresponding to the background. By removing the background cluster and merging the other two, we effectively isolated the ejecta material (highlighted in red in Fig. 9). Using the scale of 0.89 × 0.89 m² per pixel, we estimate that the ejecta covers a total area of 734036 m².

Next, we applied the *k-means* clustering algorithm again on the previously isolated ejecta pixels, exploring values of $k$ from 2 to 13, and generated color-coded maps for each solution. We found that the optimal number of clusters was 2, which are represented in green and blue (see Fig. A4). Higher $k$ values introduced artificial subdivisions that lacked spatial coherence.

To quantify the directional distribution of each cluster, we computed the azimuthal angle of each pixel relative to the crater center (measured counterclockwise from East to North). We then constructed an angular histogram (with 2° bin size) representing the pixel distribution around the crater and visualized it using polar plots (shown in the lower left and upper right corners of Fig. A4). We found that the green cluster corresponds to the central region of the crater—presumably where the ejecta layer is thickest—extending up to the innermost 50 meters and accounting for <1% of the ejecta. In contrast, the blue cluster, which represent more than 99% of the ejecta, encompasses the more distant and morphologically asymmetric part of the ejecta blanket. The distribution exhibits a pronounced asymmetry aligned along the 135°–315° azimuthal axis, indicating a preferential direction in the dispersion of the ejecta.

Based on our measured crater depth of 4.9 m, and under the approximation of a paraboloid cavity, we estimate the crater volume ($\frac{1}{2}\pi(\frac{D_{ap}}{2})^2 d$) to be approximately 1509 m$^3$. Because this excavated volume corresponds to the material redistributed as ejecta—the volume of the impactor is negligible in this case (~1 m$^3$) and a significant portion is likely vaporized upon impact (Melosh, 1989)—if we assume that it is uniformly deposited, we obtain an average ejecta thickness of approximately 2 mm.

On the other hand, McGetchin et al., (1973) derived a power-law relationship describing the decay of ejecta blanket thickness with distance from the crater (Eq. 4):

$$t = 0.14\, R^{0.74} \left(\frac{r}{R}\right)^{-3} \qquad (4)$$

where $r$ is the distance to the crater and $R$ is the crater radius in meters. This expression assumes an idealized impact with a radially symmetrical ejecta distribution and does not account for the presence of rays or the influence of local topography. Applying this relationship to the radial green cluster, which accounts for 1% of the ejecta, we find that at 50 m it would have a thickness of 49.9 mm.

These estimates aligns with findings by Housen and Holsapple (2011) who showed that most high-velocity impacts produce ejecta consists of micron- to millimeter-sized particles. Although our estimates are approximate and based on simplified assumptions, they indicate that the ejecta blanket is extremely thin—sub-centimeter in thickness—and composed of fine particulate material barely forming a continuous coating over the surface.

### *3.2 Color Changes*

As we did for the NAC images, given that the ejecta is more clearly visible at lower incidence angles, we examined all available WAC images to find an optimal balance between higher spatial resolution and lower incidence angles, while also minimizing viewing geometry differences between the pre- and post-impact datasets. We identified a dataset consisting of three pre-impact and three post-impact images (Table 2), with spatial resolutions ranging from 58 to 78 meters per pixel and incidence angles between 12° and 15°.

The new crater formed by the impact measures approximately 35 meters in diameter, meaning it spans less than a single pixel in the WAC images and therefore cannot be resolved or unambiguously identified. In fact, we generated RGB composites by assigning color filters to different bands, but no clear visual differences were observed between the pre- and post-impact images that could be confidently distinguished from the inherent noise in the dataset. To overcome this limitation, we co-aligned the WAC images with the higher-resolution NAC images by identifying larger craters and features common to both datasets and used the NAC data to pinpoint the exact pixels corresponding to the crater and its ejecta in the WAC frames.

Although the images share similar photometric angles, the wide field of view (FoV) of the WAC introduces significant variations across the scene. Therefore, photometric corrections were applied (see Fig. A5). Next, we defined three regions of interest (ROIs) to explore the area in greater detail. First, we defined a control ROI covering several hundred pixels (the number depends on the resolution of each dataset), associated with a 2 km diameter crater located about 20 km north of the new crater (blue square in the left panel of Fig. 4). This region is useful for verifying whether any artifacts in the dataset could hinder a comparative analysis. In addition, we defined what we refer to as ROI 1, spanning between 20 and 60 WAC pixels, which corresponds to the previously identified area containing ejecta. However, as noted earlier, the ejecta is not uniformly distributed and shows clear gaps. To isolate a more representative portion of the ejecta, we defined the ROI 2, which covers just 2 to 4 WAC pixels from the central ejecta region (see Fig. 4, right panel: the irregularly shaped spot corresponds to ROI 1, while the red square marks ROI 2).

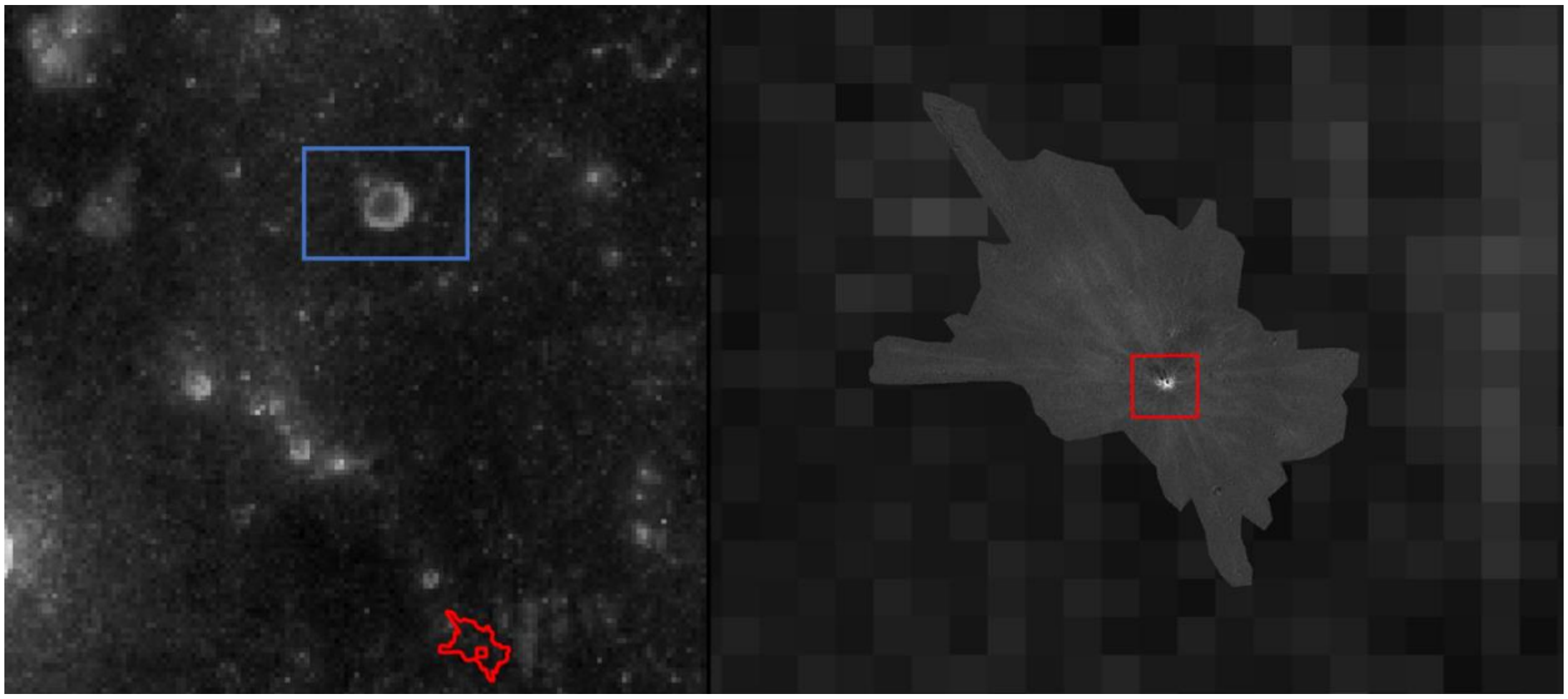

**Fig. 4. Left panel: WAC image M170695002CE, 689 nm filter, photometrically corrected. At the top, the crater selected as the control point is marked with a blue box. At the bottom, ROIs 1 and 2 are shown. Right panel: Close-up view of the ejecta blanket. The irregular dark patch corresponds to ROI 1. ROI 2 is confined to the area outlined by the red square.**

We then extracted spectra from the entire frames—referred to as the global scene—as well as from the control ROI, ROI 1, and ROI 2. This was done for each pair of images selected for this color analysis (Table 2), and the spectra were plotted together. An example is shown in Fig. 5. Additional examples, combining different pre- and post-impact datasets, are provided in the appendix (Fig. A6). In all cases, the spectra from the global scene, the control ROI, and ROI 1 are remarkably similar. By contrast, ROI 2 consistently exhibits a higher reflectance in the visible range, while the UV bands remain largely unaffected—without exception across all datasets. Consequently, when the spectral slope is measured from 321 nm, the increased contrast between the UV and visible bands manifests as a reddening trend.

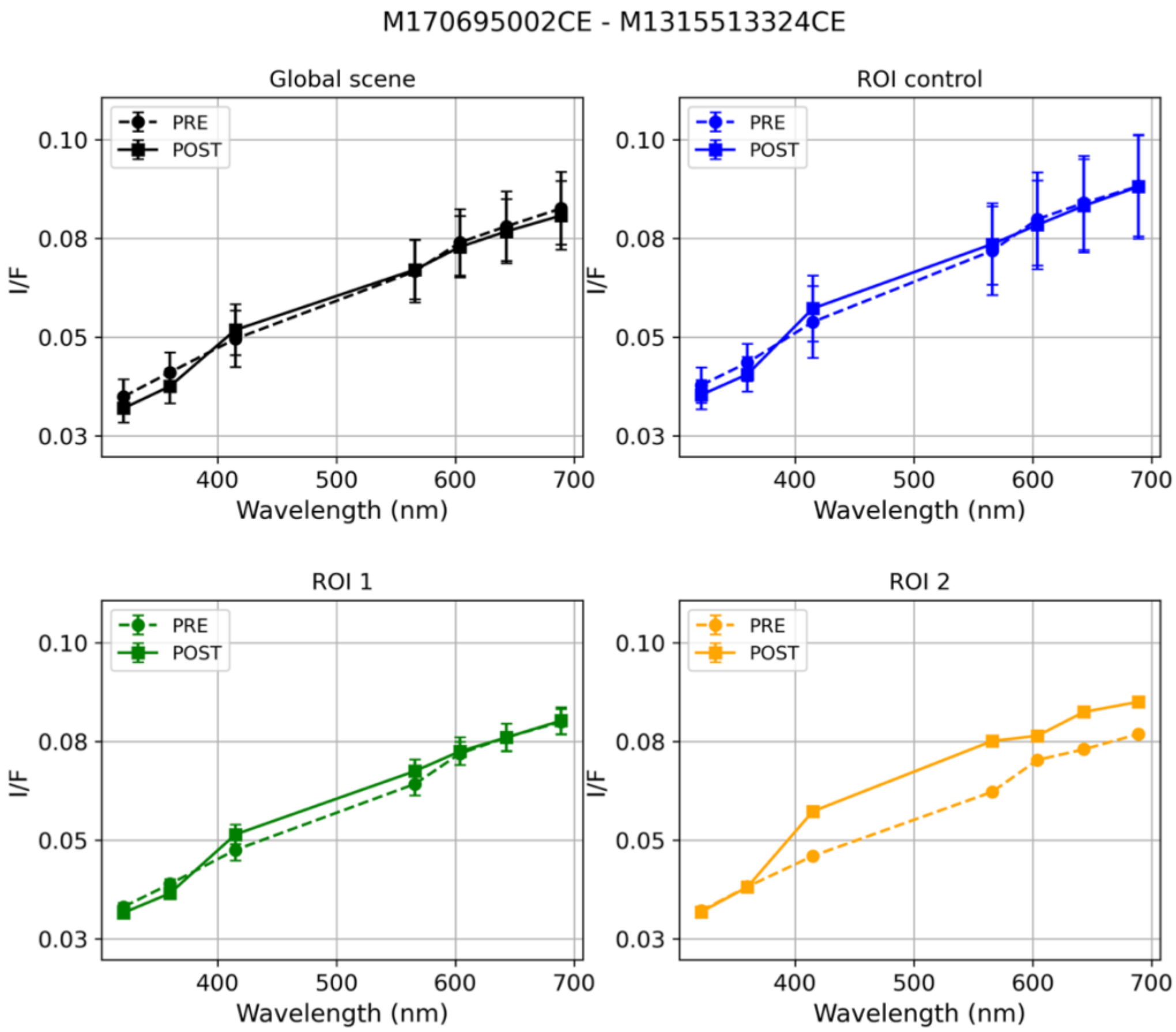

**Fig. 5. Pre- and post-impact spectra from WAC datasets M170695002CE and M1315513324C, extracted from the global scene, the control ROI, and ROIs 1 and 2. In ROI 2, a spectral reddening is evident when the slope is measured starting at 321 nm, due to enhanced reflectance in the visible range. Error bars represent the standard deviation of the data for each case.**

To further emphasize the differences, we calculated the spectral slope between all band pairs for each spectrum and normalized it by the corresponding slope from the global scene. This approach ensures that any systematic bias present in the dataset is effectively removed.

We found that the band slopes that best capture the variations are 415/321 nm, 604/321 nm, and 643/321 nm. In Fig. 6, we show in red the spectral slopes normalized to the global scene for each ROI before the impact, with the box size indicating the standard deviation. In blue, we plot the same normalized spectral slopes after the impact together with their associated uncertainty.

ROI 1 shows signs of reddening in some instances, whereas ROI 2 consistently exhibits clear reddening across all cases, corresponding to the area with the highest concentration of ejecta. Although the signal in all cases remains within 1σ and therefore lacks formal statistical significance, the consistent appearance of this trend strongly suggests a real color change following the impact. Additional examples combining different pre- and post-impact datasets are shown in the appendix (Fig. A7). In all of them, spectral reddening is consistently observed in ROI 2, in some cases reaching 1σ significance.

The largest difference is observed when comparing the spectral slopes between the 643 nm and 321 nm color filters before and after the impact. For the specific case of the ROI2 shown in Fig. 6, the slope increases by 19.3%. We performed this calculation for all filter combinations—including those presented in Fig. A7—and found an average spectral reddening of 16.54% between the pre- and post-impact spectra for the 643 nm/321 nm ratio.

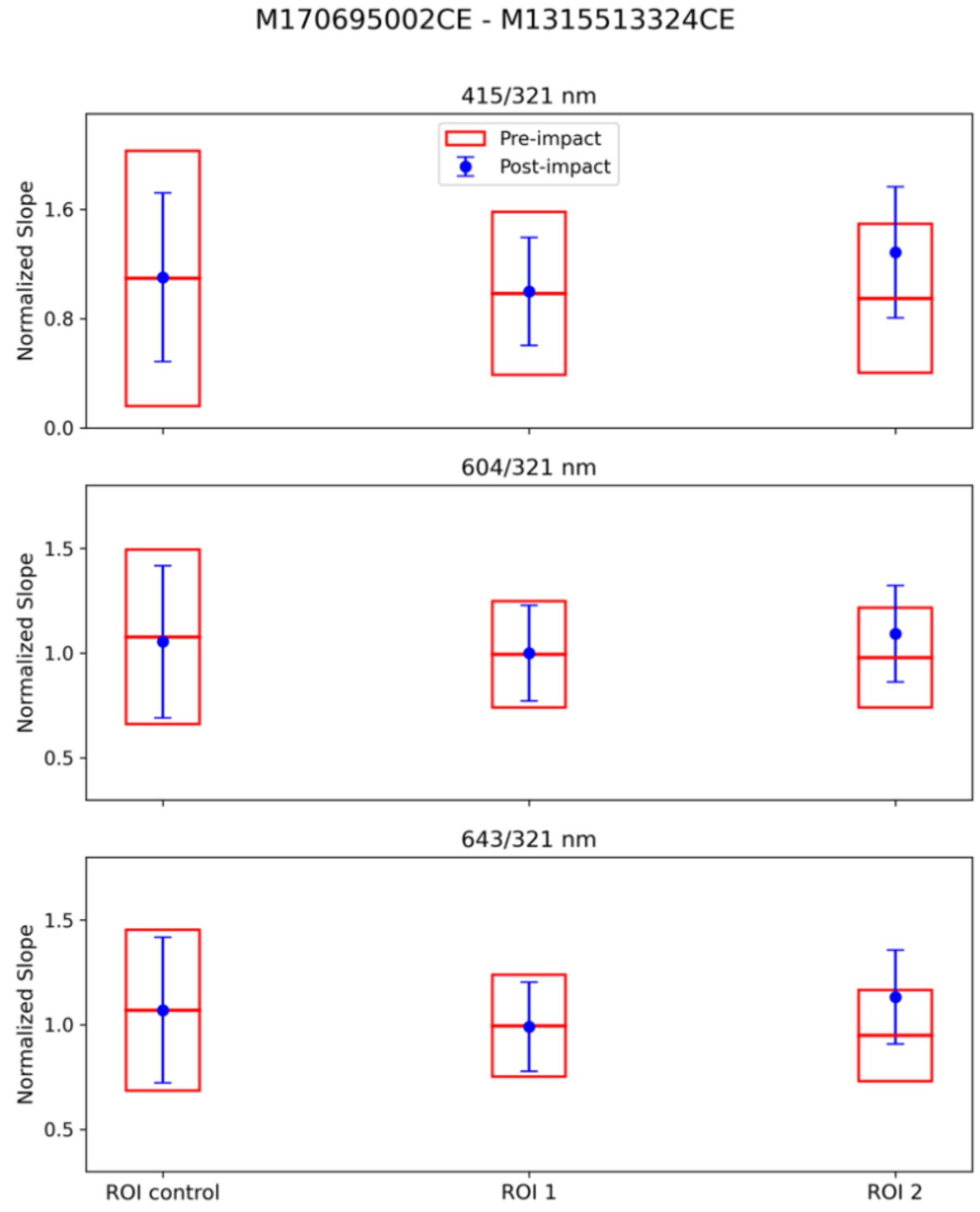

**Fig. 6. Spectral slopes normalized to the global scene (415/321, 604/321, and 643/321 nm) for the pre- and post-impact WAC datasets M170695002CE and M1315513324C, for the control ROI, ROI 1, and ROI 2. Red bars**

**show the normalized spectral slope before and blue bars after the impact. ROI 2 exhibits a noticeable reddening, with the 643/321 nm ratio showing the most pronounced change—an increase of 19.3%, approaching the 1σ level.**

## 4. Discussion

According to Melosh (1989), the ejecta morphology becomes increasingly asymmetric as the impact angle measured from the horizon decreases:

- For angles < 90°, ejecta is preferentially concentrated on the downrange side, i.e. the region corresponding to the direction and sense of the projectile's trajectory.
- At < 45°, a wedge-shaped "forbidden zone" appears uprange, where little to no ejecta is deposited.
- At < 20°, the previous pattern becomes more pronounced, appearing now two forbidden regions, downrange and uprange, and forming a distinct butterfly-shaped ejecta blanket.

The observed morphology, with the presence of two forbidden zones and the overall butterfly-shaped pattern, suggests an oblique impact with an elevation angle below 20°, though not lower than 10°, as angles below this threshold typically result in a loss of crater circularity (Melosh 1989)—which is not observed in our case. It is also supported by the numerical simulations for low-size impactors striking at an angle of <20°, according to Luo et al., (2022). As shown in Fig 6b of this work (impact velocity of 15 km/s), the simulations reveal distinctive wing-shaped structures at the outer edges of the ejecta blanket—features that are also visible in our dataset.

Therefore, the evidence strongly points to a most plausible impact direction that is perpendicular to the dominant axis of the ejecta distribution—that is, forming an azimuthal angle of approximately 45° or 225° (measured counterclockwise from east, see green dashed line in Fig. 9), and with an elevation angle between 10° and 20°.

To quantitatively evaluate theses impact elevation angles, we turn to analytical formulations that relate impact geometry to crater dimensions. By the late 20th century, there was a surge of interest in understanding the physics underlying crater impacts and establishing relationships between crater diameter and impact parameters (e.g. Öpik, 1969; Gault, 1974; Croft, 1977; Dence et al., 1977; Shoemaker and Wolfe, 1982; Holsapple and Schmidt, 1982; Melosh, 1989, among others). One such example is the equation proposed by Öpik (1969):

$$D = 9.24 \cdot 10^{-3}\, k^{1/2}\, M^{1/3}\, V^{0.467}\, \rho_p^{-1/12} \cos(\theta_n)^{-1/2} \qquad (5)$$

given in CGS units, with $k$ is obtained by solving the following equation

$$k^2(1 + 0.04c^2V^4) - 4k + 4 - c^2V^4 = 0$$

$$2 \leq k \leq 5, \quad c = 4.2 \times 10^{-13}\ \mathrm{s^2/cm^2}$$

with $D$ the final crater diameter, $M$ the mass of the impactor, $V$ the impact velocity, $\rho_p$ the bulk density of the impactor, and $\theta_n$ the meteoroid incidence angle with respect to the surface normal[10].

Another widely used equation that includes a dependence on the impact angle is that of Gault, (1974), valid for craters up to 100 meters in diameter in weakly cohesive particulate material similar to the lunar regolith, as it is the case:

$$D_{ap} = 0.25\,\rho_p^{1/6}\,\rho_t^{-1/2}\,E^{0.29}\,(\sin\theta_t)^{1/3} \qquad (6)$$

given in MKS units, where $D_{ap}$ is the apparent diameter, with $\rho_t$ the density of the target (Moon) and $E$ the kinetic energy of the impactor. Note that, as discussed in Section 3.1, rim-to-rim diameters (*D)* are approximately 25% larger than the apparent diameters ($D_{ap}$).

Finally, it can be derived from the experimental results of Schmidt and Housen (1987) (Dry Sand, $\Pi_R = 0.84\Pi_2^{-0.17}$) the following equation[11]:

$$D_{ap} = 1.49\,\rho_p^{0.17/3}\rho_t^{-1/3}M^{0.83/3}(V\sin\theta_t)^{0.34}g^{-0.17} \qquad (7)$$

given in MKS units, where $D_{ap}$ is the apparent diameter and $g$ is the lunar gravitational acceleration, which has a value of 1.62 m/s². Note that Although it does not appear in the original formulation of Schmidt and Housen (1987), we replaced $V$ with $V\sin\theta_t$ to account for the dependence on the impact angle.. These scaling laws highlight the complex interplay between impactor properties, target characteristics, and impact geometry in shaping final crater dimensions. However, they differ significantly in both their derivation and formulation—ranging from semi-empirical fits to dimensional analysis and experimental scaling. Given these differences, we proceed to examine which of these formulations best reproduces the observed crater. We first analyze the angular dependence of the three expressions: Öpik (1969), Gault, (1974), and Schmidt and Housen (1987). To isolate the effect of the impact angle, we set all multiplicative factors to unity and plotted the resulting crater diameter as a function of impact angle (Fig. 7). We find that the Öpik formulation predicts a divergence in crater diameter for tangential impacts, tending toward infinity—a counterintuitive result that was already noted and criticized by Gault, 1974. For this reason, we exclude the Öpik equation from this work.

[10] Note that other authors define the impact angle relative to the horizontal (i.e., the elevation angle). To avoid confusion, throughout this work we adopt the notation $\theta_n$ when referring to the normal-referenced angle, measured from the vertical, and $\theta_t$ when referring to the tangential-referenced angle, measured from the horizontal.

[11] Note that we replaced $V$ with $V\sin\theta_t$ to account for the dependence on the impact angle, which is not included in the original formulation of Schmidt and Housen (1987).

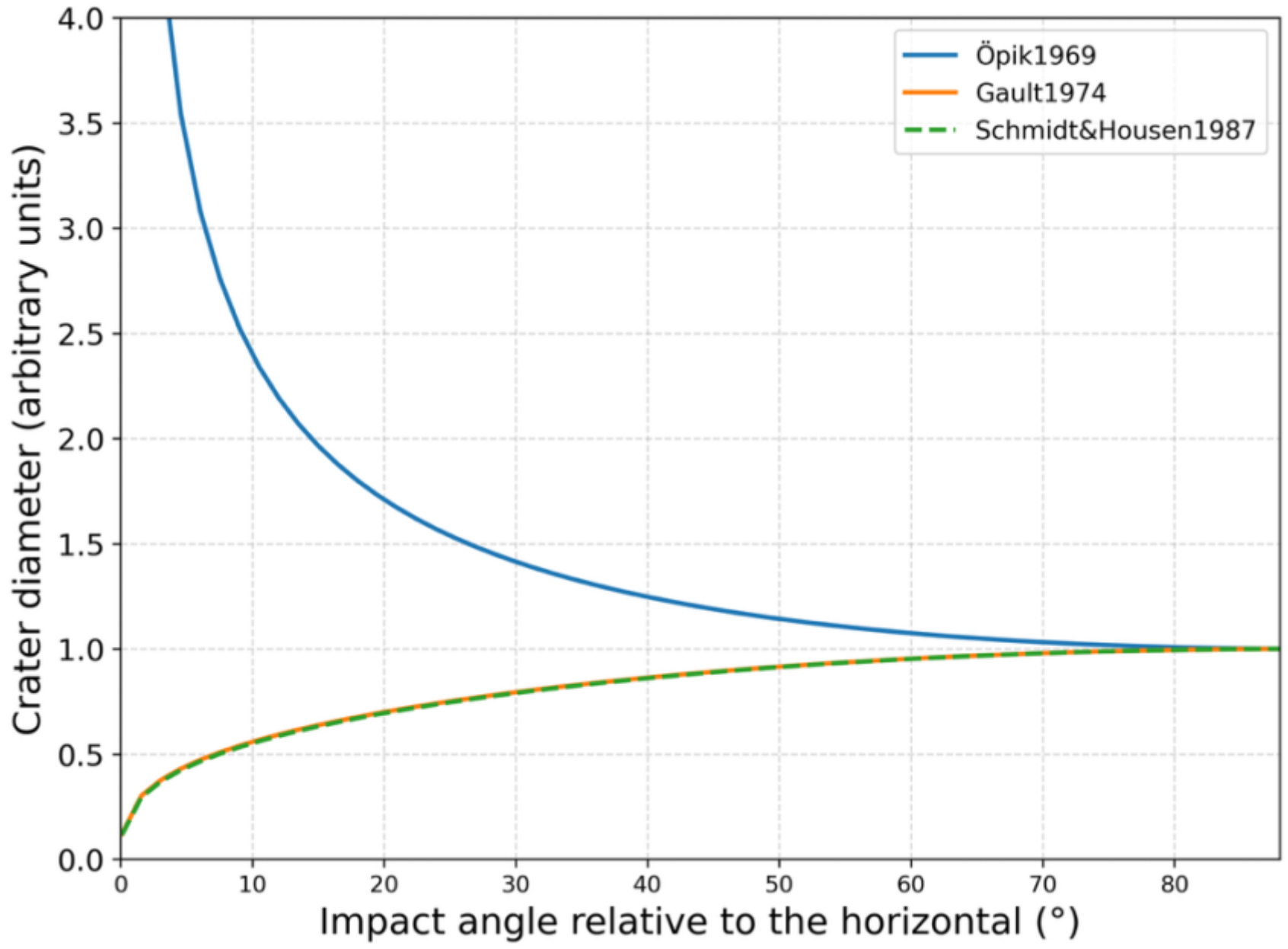

**Fig. 7. Angular dependence of crater diameter as predicted by three commonly used scaling laws: Öpik (1969), Gault (1974), and Schmidt & Housen (1987). To isolate the influence of impact angle, all constant terms were normalized to unity, and the crater diameter is plotted in arbitrary units. The Öpik formulation predicts a diverging diameter for tangential impacts (0° relative to the horizontal). In contrast, the Gault and Schmidt & Housen models show more physically consistent trends, with smaller diameters at shallow angles and maxima near vertical incidence.**

We then evaluate the Gault (1974) and Schmidt & Housen (1987) scaling laws for our specific case by substituting the known parameters into each formulation. According to Fa (2020), the bulk density of the lunar regolith at the Chang'e-3 landing site increases with depth, ranging from approximately 0.85 g/cm³ at the surface to 2.25 g/cm³ at a depth of 5 meters. Given that the estimated depth of our crater is at least 4.9 ± 0.4 m, we confidently adopt the intermediate value of 1.6 g/cm³ proposed by Ortiz et al., (2015) for this impact analysis. Next, we use a range of bulk density of the impactor $\rho_p$ between 0.3 g/cm³, representative of cometary materials with high porosity and low compaction (Movshovitz et al., 2012) and 3.7 g/cm³, representative of planetary surfaces like that of Mars (Baratoux et al., 2014). For velocity, we use a typical impact velocity of 17 km/s (Madiedo et al. 2014). Finally, we need a luminous efficiency ($\eta$) value to convert observed energy into kinetic energy (or impactor mass), as required by the scaling laws. Madiedo et al. (2014) used a value of $\eta = 2 \times 10^{-3}$, and we adopt the same value for consistency with their analysis.

Fig. 8 shows the predicted crater diameter (rim-to-rim diameter, obtained by multiplying the apparent diameter derived from Eqs. 6 and 7 by 1.25 to account for the ~25% difference) as a function of impact angle (measured relative to the horizontal) for both scaling laws. The gray horizontal line indicates the observed rim-to-rim crater diameter (*D)* of 35 meters, and the dashed lines the ± 0.7 meters uncertainty. Neither of the two laws yields a range between 10° and 20°, as estimated from the ejecta morphology for this luminous efficiency.

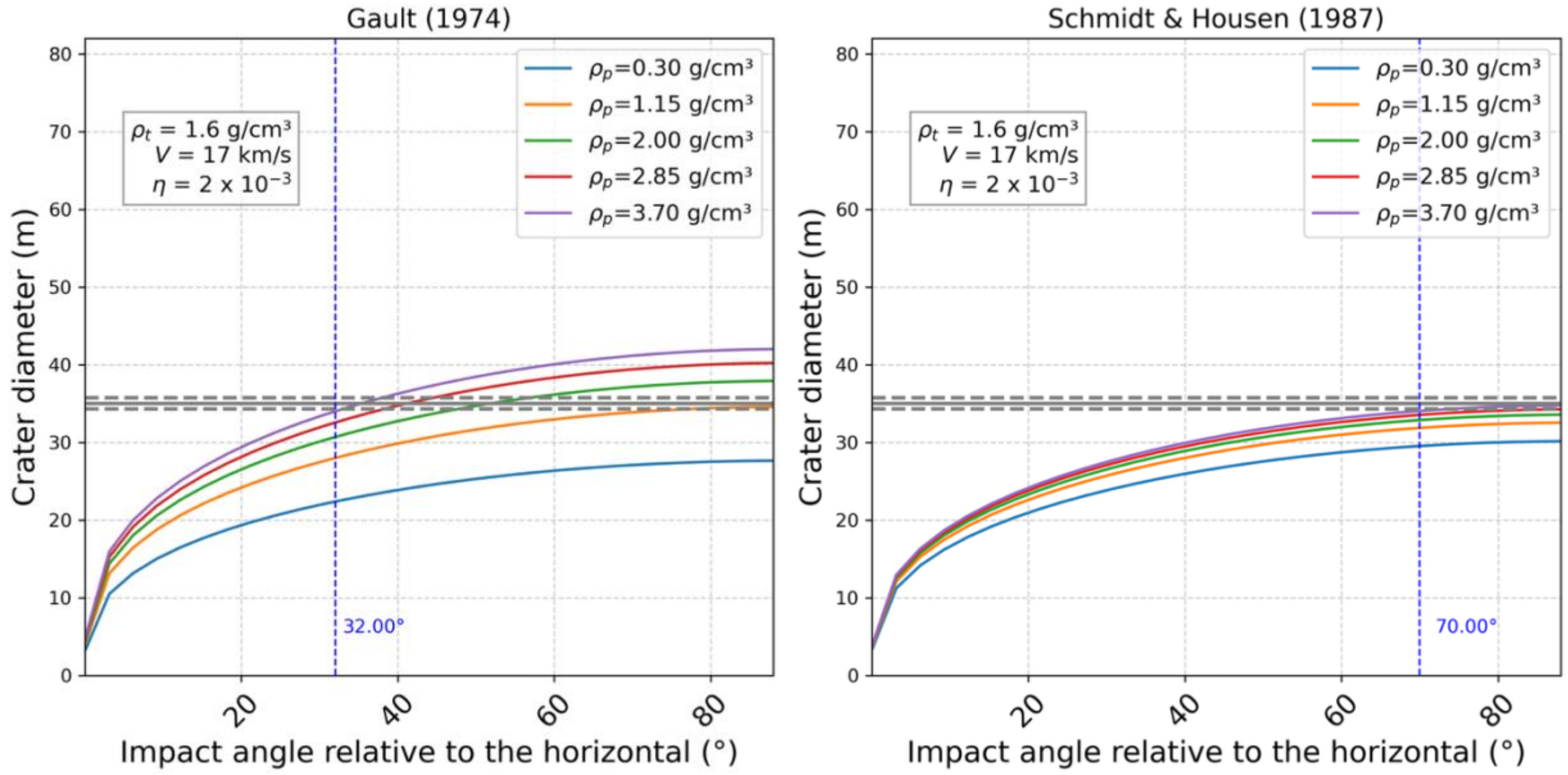

**Fig. 8. Predicted crater diameter (rim-to-rim diameter, i.e., corrected for the ~25% difference from the apparent diameter) as a function of impact angle relative to the horizontal, using the scaling laws from Gault (1974) (left) and Schmidt & Housen (1987) (right). The gray horizontal line indicates the observed crater diameter of 35 m, and dashed lines the ± 0.7 m uncertainty. Colored curves represent different impactor densities, from 0.30 to 3.70 g/cm³. Both models use the same parameters: target density $\rho_t$ = 1.6 g/cm³, impact velocity $V$ = 17 km/s, and luminous efficiency $\eta$ = 2×10⁻³. According to Gault (1974), the impact angle must be greater than 32°, whereas the Schmidt & Housen (1987) formulation requires angles above 70° (see dash blue lines).**

Swift et al., (2011) or Bouley et al., (2012) adopted a value of $\eta = 1.5 \times 10^{-3}$. Using this value, the Gault (1974) formulation yields an impact angle above 24° when using the (see Fig. A8), which is closer to the expected range. In contrast, the Schmidt & Housen (1987) expression is not compatible with the observation due to it yields over 50° (see Fig. A9). For the higher value of $\eta = 6 \times 10^{-3}$ proposed by Sheward et al., (2025), both predictions are incompatible with the scaling laws. To constrain the impact angle to lie between 10° and 20°—consistent with the ejecta morphology described by Melosh 1989 and reproduced in the simulations by Luo et al., 2022—, only the Gault, (1974) scaling law proves adequate, and it requires adopting a lower $\eta$ value and higher densities (in Figs. A8 and A9 we also explored the $7 \times 10^{-4}$ value suggested by Ortiz et al., (2015). To place density values in context, the lowest bulk densities among Solar System bodies are found in comets and primitive asteroids. For example, comet 9P/Tempel 1 has a bulk density of approximately 0.6 g/cm³ (A'Hearn et al., 2005), while the primitive asteroid Bennu has a density of about 1.2 g/cm³ (Lauretta et al., 2019). In contrast, silicaceous asteroids are significantly denser: the S-type asteroid (433) Eros has a bulk density of around 2.6 g/cm³ (Wilkison et al., 2002), and the differentiated V-type asteroid (4) Vesta exhibit densities of approximately and 3.4 g/cm³ (Russell et al., 2012).

Our analysis indicates that the Gault (1974) scaling law more accurately reproduces the observed crater size, and that luminous efficiency values below $\eta = 2 \times 10^{-3}$ and higher densities are more consistent with the measured geometry. Still, these results should be interpreted cautiously, as they merely reflect the overall behavior given the uncertainties in assumed impact velocity, lunar regolith density (variable with depth), impactor density, and the poorly constrained luminous efficiency.

Madiedo et al. (2014) suggested that, given the date of the event, the impactor could have been associated with the September ε-Perseids meteor shower (SPE) originated from the comet 109P/Swift-Tuttle (geocentric radiant $RA$ = 47.67° ± 0.04°, $DEC$ = 39.493° ± 0.013°, $V$= 64.79 ± 0.06 km/s relative to Earth (Shrbený and Spurný, 2019). To investigate this possibility, we calculated the selenocentric radiant of the SPE for the Moon considering the relative motion to the Earth at the time of the impact. This transformation yields a slightly shifted radiant: $RA$ = 47.69° ± 0.04°, and $DEC$ = 39.520° ± 0.013°, which correspond to an azimuth of 80.56° and an elevation angle $\theta_t$ of 51.39° at the impact location (see yellow arrow in Fig. 9).

To establish an upper bound on the impact angle, we assume the elevation of 20° consistent both with the qualitative description of the ejecta morphology by Melosh 1989 and with the simulations presented by Luo et al., (2022). Using Eq. 8, where $\psi$ is the azimuth and $\theta_t$ is the elevation angle, we calculate the angular separation, $\xi$, between this trajectory and the SPE radiant, obtaining a value exceeding 40°. Even when adopting higher elevation angles, such as the 32° limit obtained using the Gault (1974) expression for $\eta = 2 \times 10^{-3}$ and $\rho_p = 3.7$ g/cm³, the computed angular separation from the SPE radiant is 32.4°. This large deviation suggests that this meteor shower is unlikely to be the source of the impactor.

$$\xi = \cos^{-1}\left[\sin\theta_{t_A}\sin\theta_{t_B} + \cos\theta_{t_A}\cos\theta_{t_B}\cos(\psi_A - \psi_B)\right] \qquad (8)$$

We finally estimate the probability that the impactor belonged to the SPE by employing the technique developed by Madiedo et al., (2015), which links meteoroid streams with lunar impact flashes. This probability, $p^{SPE}$, can be quantified as:

$$p^{SPE} = \frac{\nu^{SPE}\,\gamma^{SPE}\cos(\theta_n)\;\sigma\;ZHR^{SPE}_{Earth}}{\nu^{SPO}\,\gamma^{SPO}\;HR^{SPO}_{Earth} + \nu^{SPE}\gamma^{SPE}\;\cos(\theta_n)\;\sigma\;ZHR^{SPE}_{Earth}} \qquad (9)$$

where $\sigma$ can be approximated as 1.0 (Madiedo et al., 2015), the value $ZHR^{SPE}_{Earth}$ is 5 meteors per hour (McBeath, 2012) and $HR^{SPO}_{Earth}$ is 10 meteors per hour (Dubietis and Arlt, 2010). The gravitational focusing factors $\gamma^{SPE}$ and $\gamma^{SPO}$ are 0.97 and 0.71, respectively.

In Eq. 9, the factors $\nu^{SPE}$ and $\nu^{SPO}$ are calculated as (Bellot Rubio et al., 2000):

$$\nu = \left(\frac{m_0 V^2}{2}\right)^{s-1} E_{\min}^{1-s} \qquad (10)$$

Here, $m_0$ is the mass of a shower meteoroid that would produce, on Earth, a meteor of magnitude +6.5. This corresponds to $4.9 \cdot 10^{-6}$ kg for the SPO and $3.3 \cdot 10^{-8}$ kg for the SPE stream, according to equations (1) and (2) in Hughes (1987). $V$ is the impact velocity; for SPO it is 17 km/s (Ortiz et al., 1999) while for the SPE it is 64.31 km/s (after correcting the geocentric 64.79 km/s (Shrbený and Spurný, 2019) to the selenocentric velocity). $E_{min}$ is the minimum radiated energy detectable on the Moon using the telescopes employed, as observed from Earth. For our case $E_{min}$= 3.1·10$^{6}$ J, corresponding to a limiting lunar flash magnitude of 10, according to equations (17) and (18) in Madiedo et al., (2015). Finally, we adopt for SPE a population index $r$ of 3, based on McBeath (2012), which implies a mass index $s$ = 2.19 according to equation (13) in Madiedo et al. (2015). This yields $\nu^{SPE}$ = 2.78 · 10$^{-6}$ and $\nu^{SPO}$ = 4.54 · 10$^{-5}$.

By substituting the meteoroid incident angle $\theta_n$ = 38.61 (90° − 51.39°) for the SPE into Eq. (9), we obtain a probability of 3.2% that the event was caused by an SPE meteoroid. Or in other words, there is over a 96% probability that it was a sporadic impactor (SPO). This result is fully consistent with the findings discussed above.

Our conclusion is consistent with the findings of Madiedo et al. (2014), who noted that although the September ε-Perseids (SPE) meteoroid stream experienced an outburst on September 9, its activity had returned to normal levels by September 11—the date of the lunar impact. Moreover, the estimated mass of the lunar impactor significantly exceeded that of the SPE meteoroids responsible for the brightest meteors observed on Earth during the outburst, further supporting a sporadic, unrelated origin.

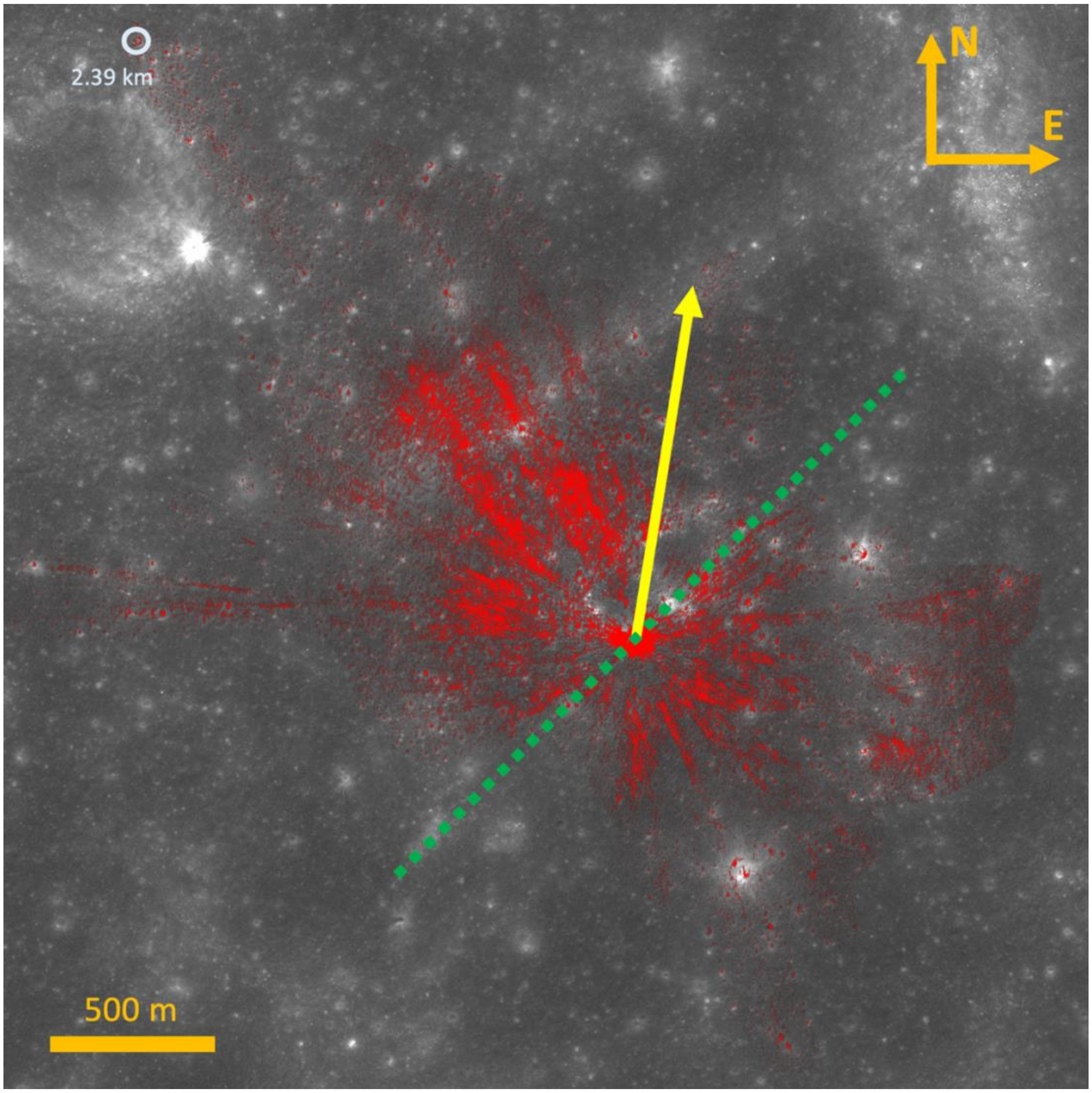

**Fig. 9. Isolated ejecta identified through clustering (red pixels) overlaid on the post-impact image M1417641871LE. The most distant ejecta lies over 2 km from the crater (grey circle in the upper left corner). The green dashed line marks the most probable impact direction based on our analysis of the ejecta morphology, with an azimuth of 45° or 225° (measured counterclockwise from east). The yellow arrow indicates the incoming direction of a meteoroid if it belonged to the ε-Perseids, with an azimuth of 80.56°. This origin is ruled out, confirming the sporadic nature of the impact event.**

## 5. Conclusions

Our morphological analysis reveals that the crater measures 37 ± 1 m in the horizontal (west–east) direction and 33 ± 1 m in the vertical (south–north) direction (rim-to-rim diameters), yielding an average diameter of 35 ± 0.7 m and a total surface area of 962 ± 39 m². The crater depth was estimated to be 4.9 ± 0.4 m. The ejecta blanket, best observed under low solar incidence angles, extends more than 2 km from the crater and covers of approximately $7 \times 10^5$ m². It exhibits a clear asymmetry aligned along the 135°–315° azimuthal axis. Assuming that all excavated material was redistributed uniformly, we estimate an average ejecta thickness of approximately 2 mm. For the radially symmetric central region, we applied a power-law relationship, obtaining a thickness of 49.9 mm at 50 m from the crater center.

We also analyzed the ejecta morphology and studied several scaling laws. Among the tested formulations, the Gault (1974) provides the best agreement with the observed crater diameter, being more compatible for luminous efficiency values of $2 \cdot 10^{-3}$ or lower and higher projectile densities. Based on the asymmetric ejecta distribution, we infer a most probable impact direction with an azimuth of 45° or 225° (measured counterclockwise from east) and an elevation angle between 10° and 20°.

We computed that this trajectory lies more than 30° away from the selenocentric radiant of the September ε-Perseids stream, even under the most favorable assumptions. Moreover, using the method developed by Madiedo et al. (2015), we estimated the probability that the impactor belonged to this meteoroid stream to be only 3.2%, in agreement with our analysis. We therefore conclude that the impactor responsible for the September 11, 2013, lunar flash originated from a sporadic source.

Color analysis of WAC images reveals a clear spectral reddening in the central crater region when comparing pre- and post-impact observations, with an average increase of 16.54% in spectral slope between 321 nm and 643 nm. This represents the first identification of color changes associated with a lunar impact and it could provide the basis for developing new algorithmic approaches to detect fresh impact craters. By flagging candidate impacts in WAC observations and confirming them through high-resolution NAC imaging, it becomes possible to build a statistical framework for estimating the frequency and size distribution of recent lunar impacts.

Such an approach is not only valuable for understanding the current impact environment on the Moon—crucial for the planning and protection of future lunar facilities—but also has important implications for Earth. Our planet lacks a continuous, global, and uniform monitoring network for meteoroids, making it difficult to characterize the impact flux directly. The Moon, acting as a natural witness plate, offers a unique opportunity to study the size-frequency distribution of impactors under stable and observable conditions. In this context, lunar impact monitoring can serve as a proxy for understanding the population of small near-Earth objects, helping to close existing gaps in Earth-based measurements.

## *Acknowledgments*

The authors thank Dr. Masahisa Yanagisawa for his insightful and constructive review, which helped improve the clarity and scientific rigor of this work. The authors gratefully acknowledge the NASA Lunar Reconnaissance Orbiter (LRO) mission and its team for providing well-documented and readily accessible data products, which made this research possible. J.L. Rizos, L.M. Lara, J.L. Ortiz, and J. M. Madiedo acknowledge financial support from the Severo Ochoa grant CEX2021-001131-S funded by MCIN/AEI/10.13039/501100011033. J.L. Rizos and L.M. Lara acknowledge financial support from grant PID2021-126365NB-C21.

## Data availability

The data underlying this article will be shared on reasonable request to the corresponding author.

# Appendix A

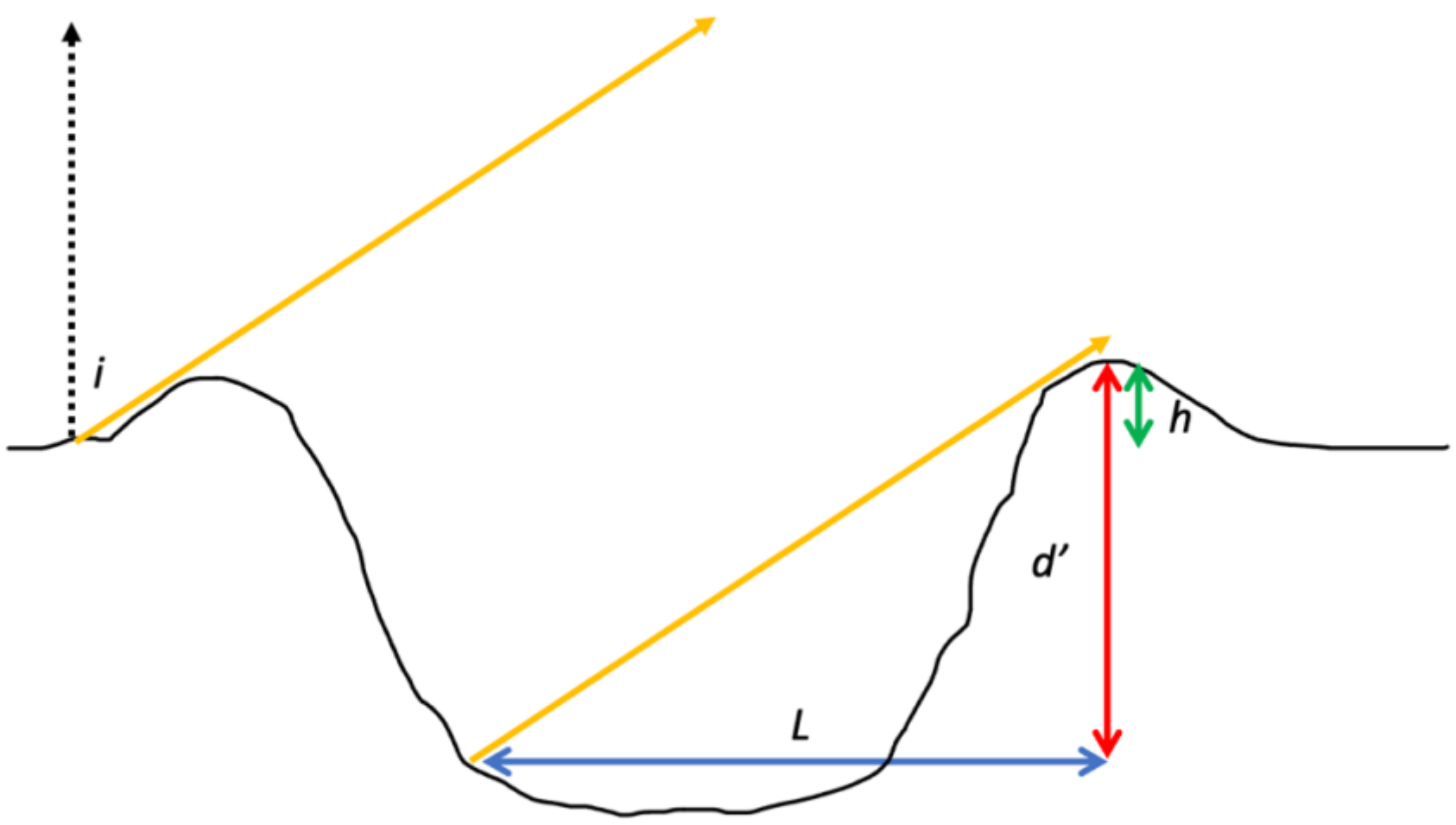

**Fig. A1. Diagram illustrating a crater casting a shadow. The yellow arrows represent the direction of incoming sunlight, while the dashed vertical arrow indicates the surface normal. The incidence angle is defined as the angle between the surface normal and the direction of the Sun. The blue arrow marks the shadow length *L*, and the red arrow indicates the depth measured from the crater rim crest to the shadow terminus *d'*. The rim crest height is indicated with the green arrow, *h*, so the crater depth *d*, is equal to *d'* minus *h* when the shadow terminus is in the center (assumed bottom) of the crater.**

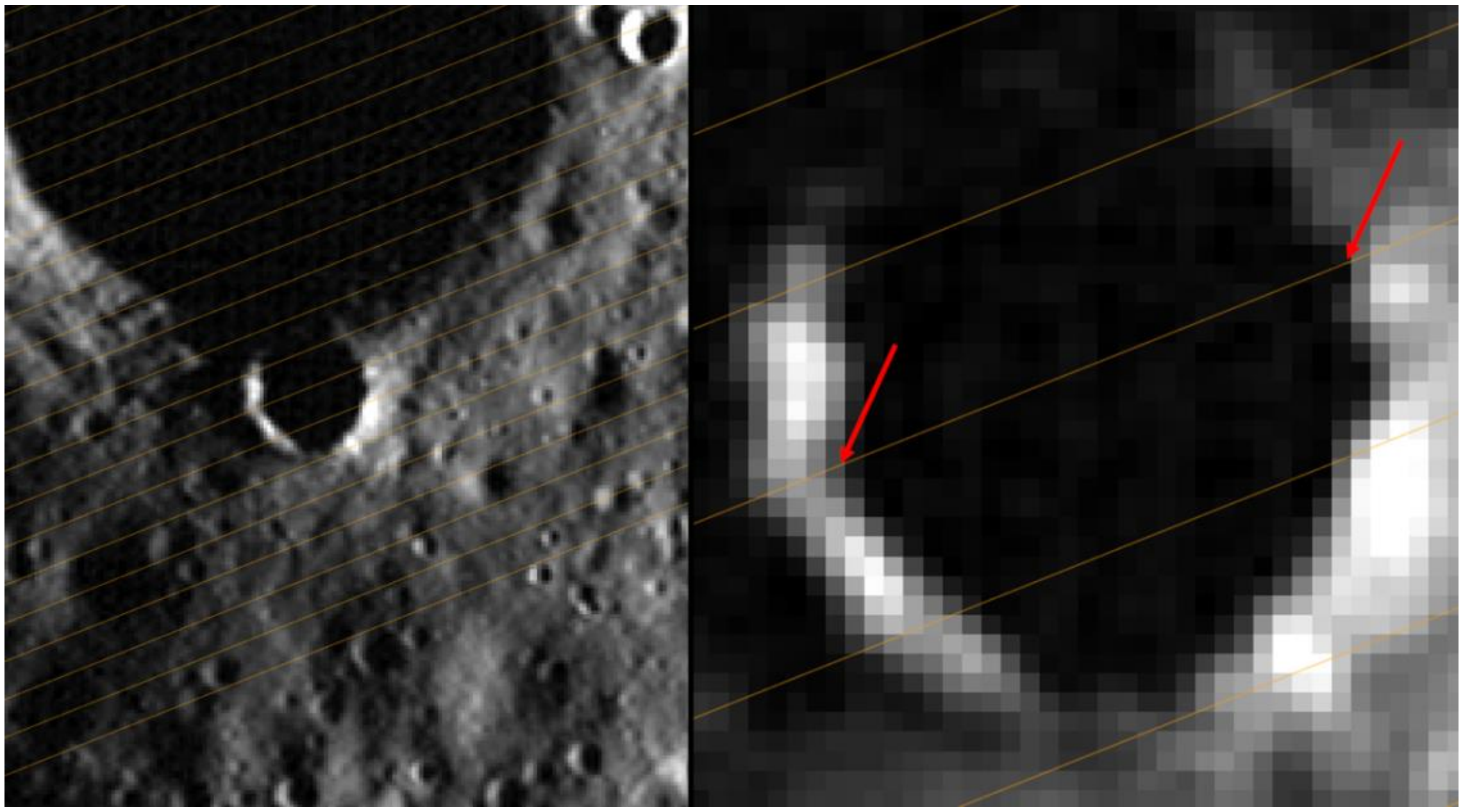

**Fig. A2. Example of estimating the minimum depth of the crater using shadow measurements. Left: NAC image M1276711033RE, with a local incidence angle of 84.44° at the crater location. Yellow lines indicate the direction of the subsolar point. Right: Zoom on the crater, where the shadow edges are measured. In this example, the shadow length is 32.5 ± 1.6 m, yielding a depth *d'* at the shadow terminus of 3.1 ± 0.2 m meters.**

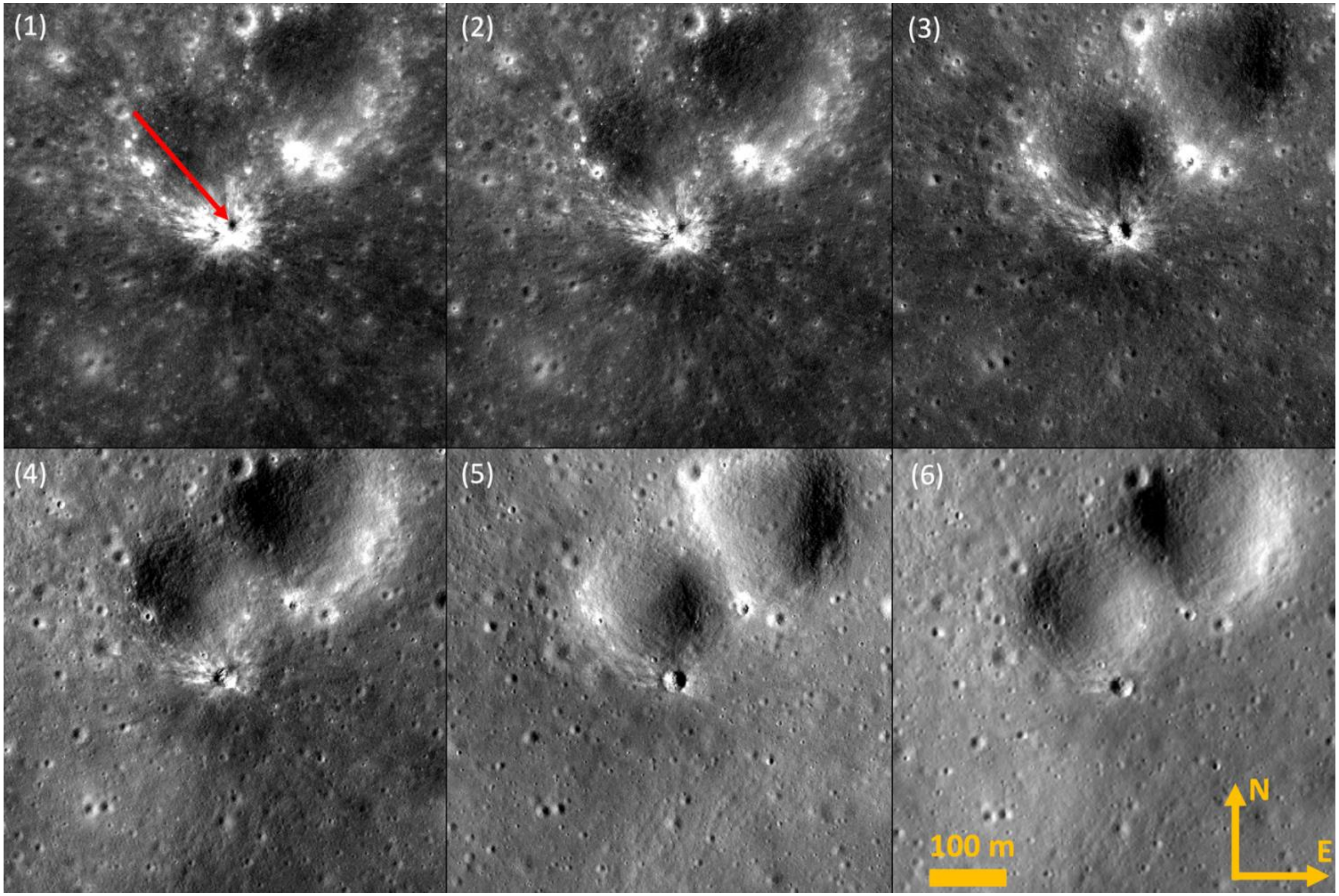

**Fig. A3. Six NAC images illustrating the visibility of the ejecta under different incidence angles. (1) M1445785486RE — incidence 22.7° (red arrow points out the new crater analyzed in this work), (2) M1371894473LE — incidence 28.9°, (3) M1435252023RE — incidence 38.8°, (4) M1471562612RE — incidence 51.0°, (5) M1320214670RE — incidence 65.6°, (6) M1441095458LE — incidence 71.1°. As the incidence angle increases, the visibility of the ejecta gradually decreases.**

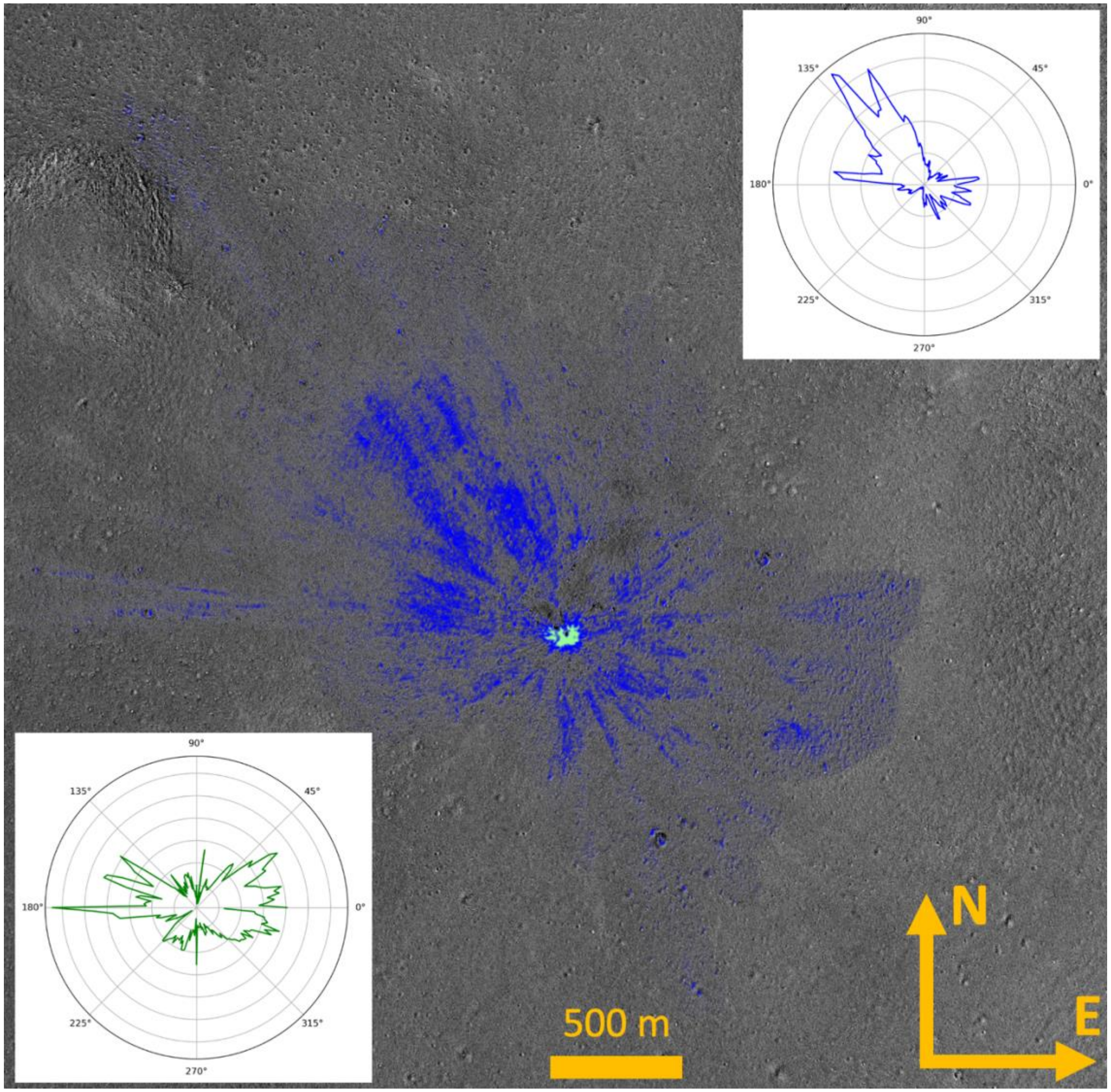

**Fig. A4. Clustering result applied to the ejecta, revealing two clusters. The green cluster, representing less than 1% of the pixels, corresponds to the crater's central region within ~50 m from the center and exhibits a circular morphology. The blue cluster, comprising over 99% of the pixels, is clearly asymmetric and aligned along the 135–315° azimuthal axis. Polar histograms in the upper and lower corners illustrate the angular distribution of each cluster using the same color code.**

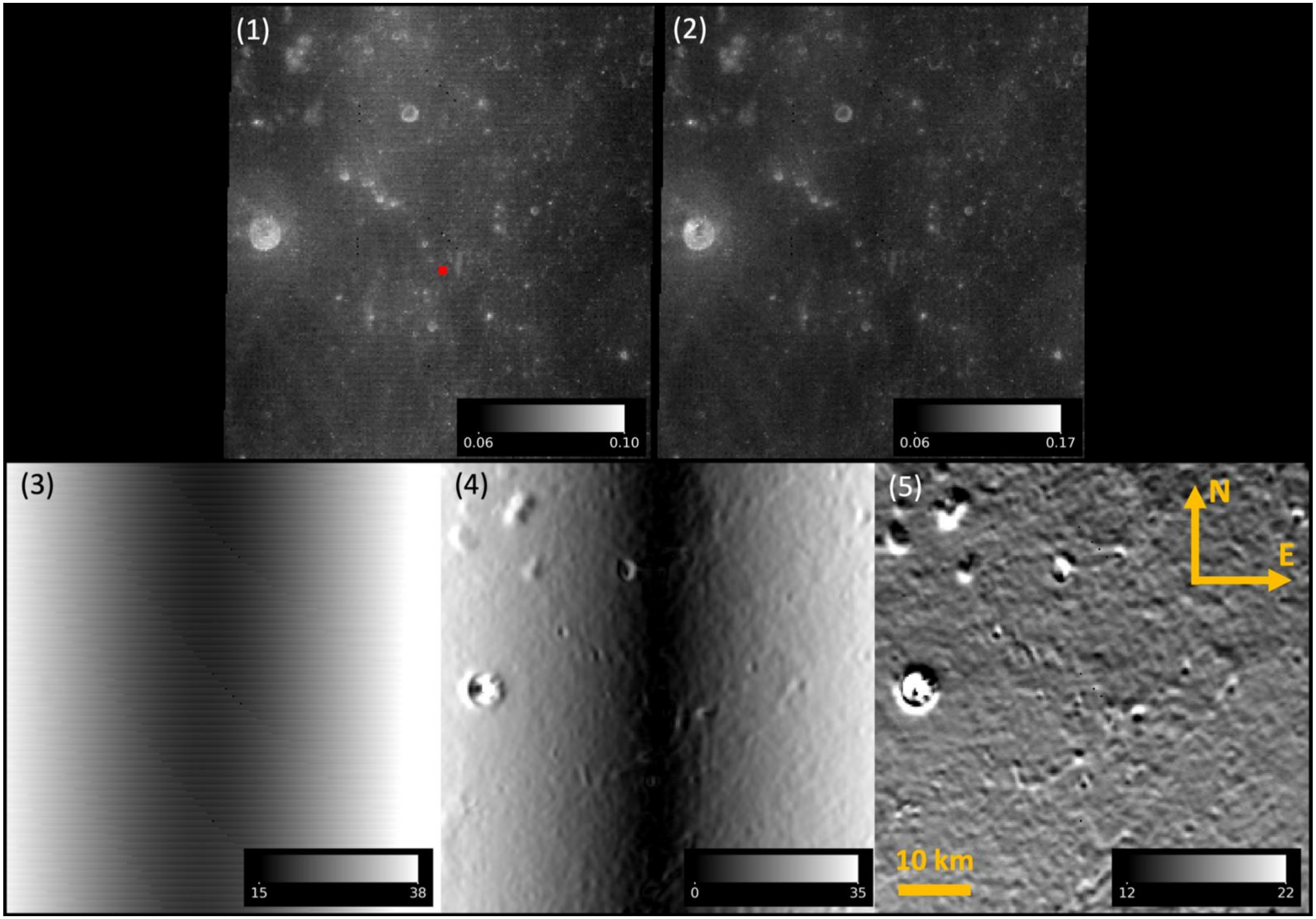

**Fig. A5.** (1) WAC color filter image at 646 nm from M1315513324CE; the red square marks the location of the newly identified crater. (2) Same as (1), but photometrically corrected. Panels (3), (4), and (5) show the phase, emission, and incidence angles corresponding to the top frames, respectively.

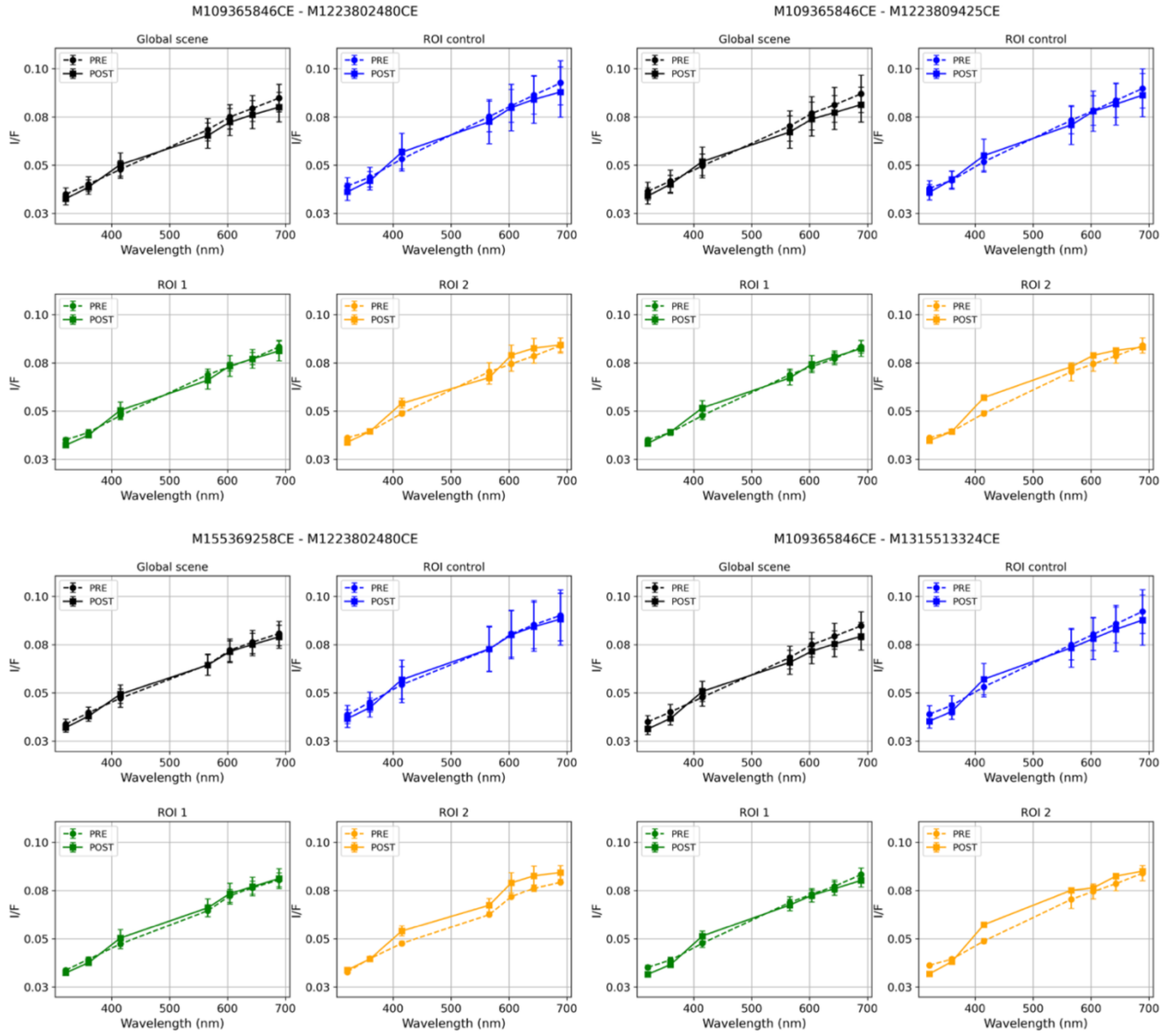

**Fig. A6. Pre- and post-impact spectra from the analysis of the global scene, the control ROI, and ROIs 1 and 2 across multiple datasets. A consistent spectral reddening from the UV to the visible range is observed in ROI 2 across all cases.**

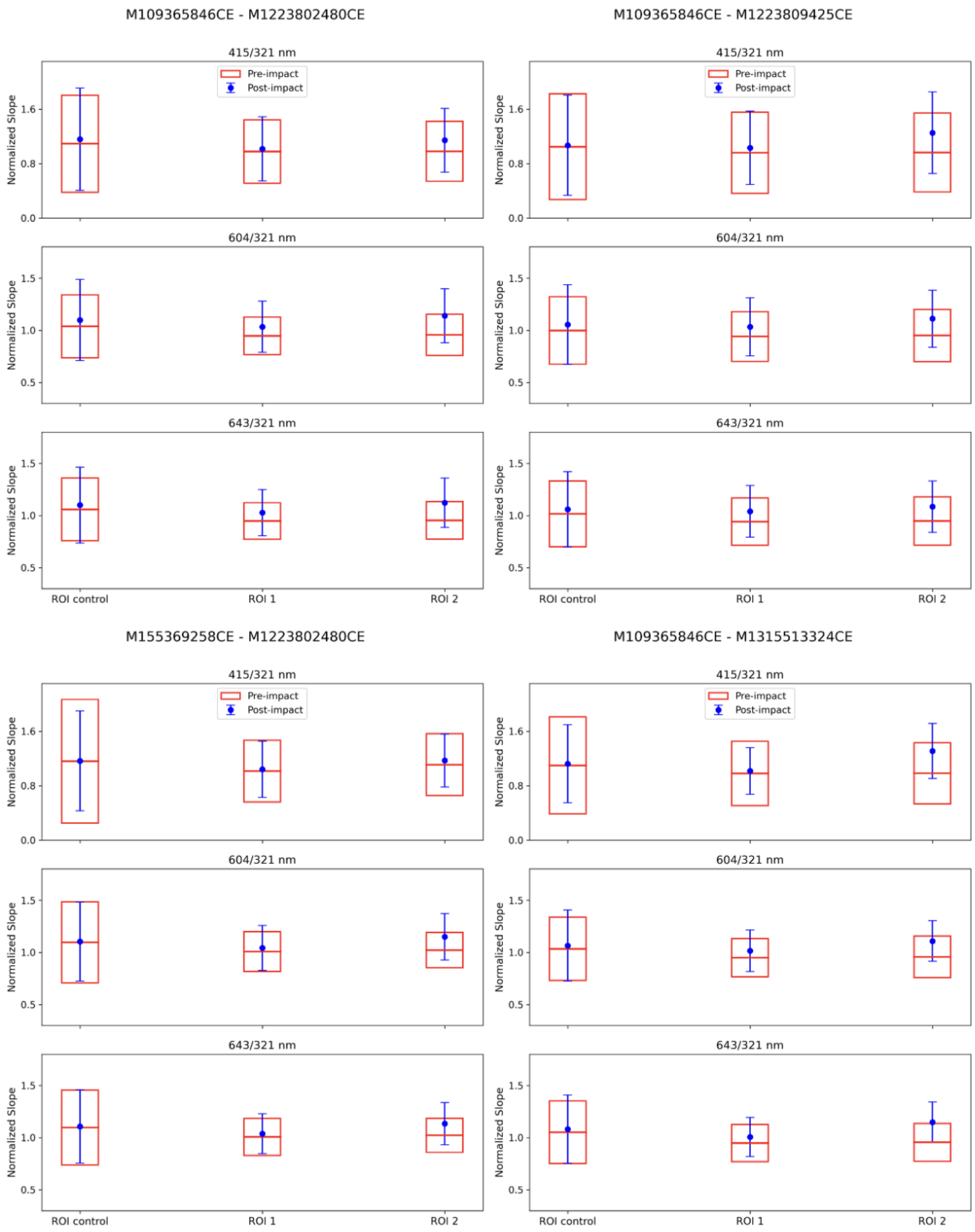

**Fig. A7. Spectral slopes normalized to the global scene (415/321, 604/321, and 643/321 nm) for the pre- and post-impact WAC datasets, evaluated for the control ROI, ROI 1, and ROI 2. Red bars indicate the normalized spectral slope in each ROI before and blue bars after the impact. ROI 2 shows a clear reddening, with ~1σ in the 643/321 nm for some case.**

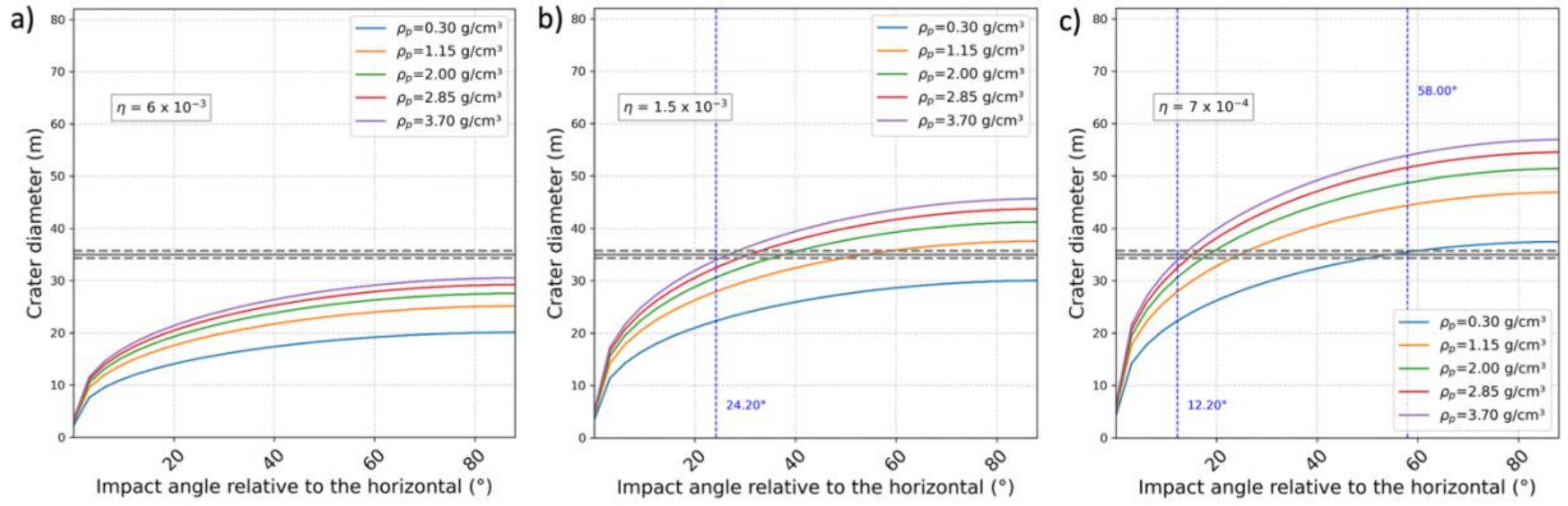

**Fig. A8. Predicted crater diameters (rim-to-rim diameter, corrected for the ~25% difference from the apparent diameter) based on the Gault (1974) scaling law (Eq. 6), which is applicable to craters up to 100 m in size formed in weakly cohesive particulate material similar to the lunar regolith. Panels (a), (b), and (c) correspond to luminous efficiency (η) values of $6\times10^{-3}$, $1.5\times10^{-3}$, and $7\times10^{-4}$, respectively. The colored curves represent a range of impactor densities from 0.3 to 3.7 g/cm³. Only the lowest value of luminous efficiency, $\eta = 7\times10^{-4}$, yield impact angles compatible with the observed butterfly-shaped ejecta pattern. In contrast, the higher value of $\eta = 6\times10^{-3}$ reported by Sheward et al. (2025), is not consistent with the measured crater diameter according to this expression.**

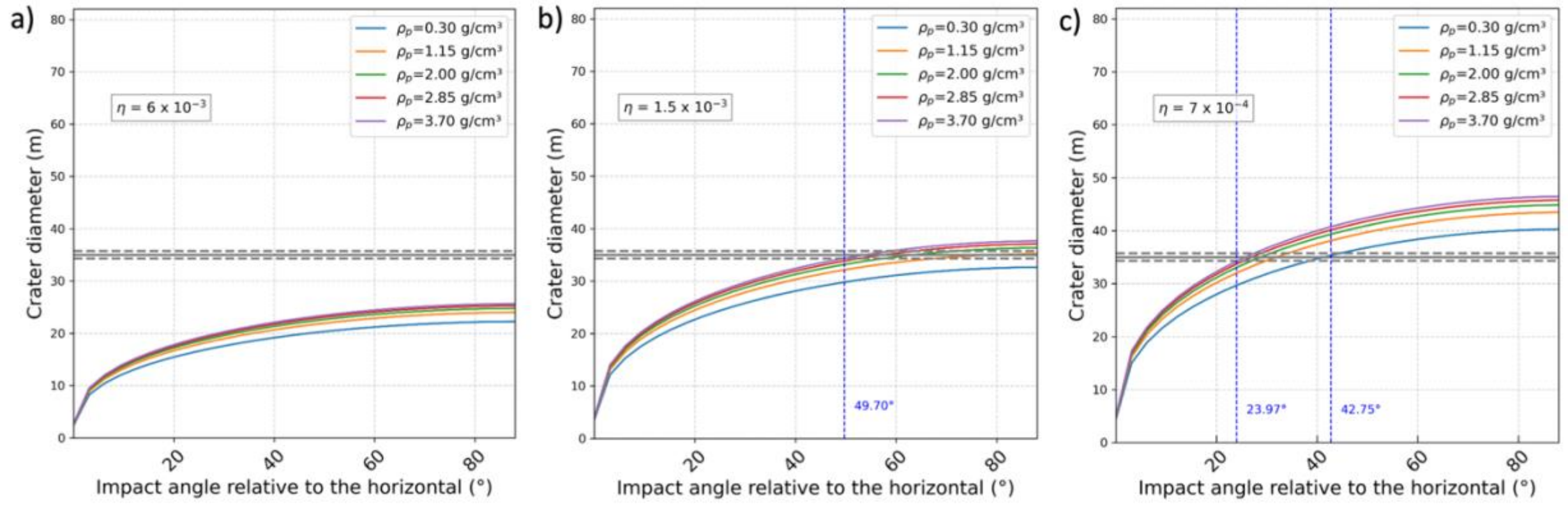

**Fig. A9. Same as Fig. A8, but using the Schmidt & Housen (1987) scaling law (Eq. 7) as a function of tangential impact angle. The two higher luminous efficiency values, $\eta = 6\times10^{-3}$ and $\eta = 1.5\times10^{-3}$, are inconsistent with the observed crater size and the morphology of the ejecta blanket. Even the lowest tested value, $\eta = 7\times10^{-4}$, requires an impact angle falling short of the <20° range suggested by Melosh (1989) and supported by the ejecta simulations of Luo et al. (2022).**